\documentclass[preprintnumbers,article,amsmath,amssymb,floatfix,10pt,prd,onecolumn,superscriptaddress,nofootinbib]{revtex4-2}

\usepackage{titlesec}
\usepackage[toc]{appendix}
\usepackage{caption}
\titleformat*{\section}{\LARGE\bfseries}
\titleformat*{\subsection}{\Large\bfseries}
\titleformat*{\subsubsection}{\large\bfseries}
\titleformat*{\paragraph}{\large\bfseries}
\titleformat*{\subparagraph}{\large\bfseries}
\usepackage{bm}
\usepackage{amsfonts}
\usepackage{latexsym}
\usepackage{graphicx}
\usepackage{amsmath}
\usepackage{palatino}
\usepackage{mathpazo}
\usepackage{textcomp}
\usepackage{float}
\usepackage{booktabs}
\usepackage{dcolumn}
\usepackage{ragged2e}
\usepackage{hyperref}
\hypersetup{colorlinks,citecolor=blue}
\hypersetup{colorlinks=true,linkcolor=blue,filecolor=magenta,    urlcolor=blue}
\usepackage{amsthm}
\usepackage{xcolor}
\usepackage{orcidlink}
\usepackage{epsfig}
\usepackage{subcaption}
\usepackage{commath}
\usepackage{cancel, ulem}
\usepackage[utf8]{inputenc}
\usepackage{multirow}
\usepackage{tablefootnote} 
\newcommand{\m}{\mathring}

\def\jnl@style{\it}
\def\aaref@jnl#1{{\jnl@style#1}}

\def\aaref@jnl#1{{\jnl@style#1}}

\def\aj{\aaref@jnl{AJ}}                   
\def\apj{\aaref@jnl{ApJ}}                 
\def\apjl{\aaref@jnl{ApJ}}                
\def\apjs{\aaref@jnl{ApJS}}               
\def\apss{\aaref@jnl{Ap\&SS}}             
\def\aap{\aaref@jnl{A\&A}}                
\def\aapr{\aaref@jnl{A\&A~Rev.}}          
\def\aaps{\aaref@jnl{A\&AS}}              
\def\mnras{\aaref@jnl{Mon.~Not.~Roy.~Astron.~Soc.}}             
\def\prd{\aaref@jnl{Phys.~Rev.~D}}        
\def\prc{\aaref@jnl{Phys.~Rev.~C}}  
\def\prl{\aaref@jnl{Phys.~Rev.~Lett.}}    
\def\qjras{\aaref@jnl{QJRAS}}             
\def\skytel{\aaref@jnl{S\&T}}             
\def\ssr{\aaref@jnl{Space~Sci.~Rev.}}     
\def\zap{\aaref@jnl{ZAp}}                 
\def\nat{\aaref@jnl{Nature}}              
\def\aplett{\aaref@jnl{Astrophys.~Lett.}} 
\def\apspr{\aaref@jnl{Astrophys.~Space~Phys.~Res.}} 
\def\physrep{\aaref@jnl{Phys.~Rep.}}      
\def\physscr{\aaref@jnl{Phys.~Scr}}       
\def\commat{\aaref@jnl{Comm.~Math.~Phys.}}              
\def\science{\aaref@jnl{Science}}               
\def\cqg{\aaref@jnl{Classical Quant.~Grav.}}            
\def\jpcs{\aaref@jnl{JPCS}}                                     
\def\ijmpd{\aaref@jnl{Int.~J.~Mod.~Phys.~D}}                    
\def\grg{\aaref@jnl{Gen.~Relat.~Gravit.}}               
\def\rpp{\aaref@jnl{Rep.~Prog.~Phys.}}          
\def\npa{\aaref@jnl{Nucl.~Phys.~A}}        
\def\lrr{\aaref@jnl{Living Rev.~Rel.}}                   
\def\jcap{\aaref@jnl{J.~Cosmology Astropart.~Phys.}}    
\def\rmp{\aaref@jnl{Rev.~Mod.~Phys.}}   
\def\epjc{\aaref@jnl{Eur.~Phys.~J.~C}} 
\def\plb{\aaref@jnl{~Phy.~Lett.~B}} 
\def\mpla{\aaref@jnl{Mod.~Phy.~Lett.~A}} 
\def\arxiv{\aaref@jnl{arxiv.org}}

\allowdisplaybreaks[1]

\begin{document}

\title{Scalar field coupled to boundary in non-metricity: a new avenue towards dark energy}
\author{Ghulam Murtaza\orcidlink{0009-0002-6086-7346}}
\email{ghulammurtaza@1utar.my}
\affiliation{Department of Mathematical and Actuarial Sciences, Universiti Tunku Abdul Rahman, Jalan Sungai Long,
43000 Cheras, Malaysia}
\author{Avik De\orcidlink{0000-0001-6475-3085}}
\email{avikde@um.edu.my}
\affiliation{Institute of Mathematical Sciences, Faculty of Science, Universiti Malaya, 50603 Kuala Lumpur, Malaysia}
\author{Tee-How Loo\orcidlink{0000-0003-4099-9843}}
\email{looth@um.edu.my}
\affiliation{Institute of Mathematical Sciences, Faculty of Science, Universiti Malaya, 50603 Kuala Lumpur, Malaysia}
\author{Andronikos Paliathanasis\orcidlink{0000-0002-9966-5517}}
\email{anpaliat@phys.uoa.gr}
\affiliation{Departamento de Matem\'{a}ticas, Universidad Cat\`{o}lica del Norte, Avda.
Angamos 0610, Casilla 1280 Antofagasta, Chile}
\affiliation{Institute of Systems Science, Durban University of Technology, Durban 4000,
South Africa}
\affiliation{Centre for Space Research, North-West University, Potchefstroom 2520, South Africa}
\affiliation{National Institute for Theoretical and Computational Sciences (NITheCS), South Africa}


\footnotetext{GM acknowledges Universiti Tunku Abdul Rahman Research Fund project IPSR/RMC/UTARRF/2023-C1/A09. AD acknowledges the support by Universiti Malaya BKP - Penyelidik Muda - (BKP119-2025-ECRG). }

\begin{abstract}
While conformal transformations in metric scalar-tensor theories recover General Relativity, this feature is notably absent in standard non-metricity-based theories. We demonstrate 
that by introducing the boundary term $C$, a non-metricity scalar-tensor theory can recover Symmetric Teleparallel Equivalent of General Relativity (STEGR) in the Einstein frame. Motivated by this, we propose a novel gravity model where a scalar field couples nonminimally to both the non-metricity scalar $Q$ and the boundary term $C$. We focus in the cosmological scenario where we present the covariant formulation and a unified autonomous system framework that 
treats generic affine-connection choices, including coincident and non-coincident gauges, on an equal footing. Our dynamical analysis across three connection branches reveals standard thermal histories and stable de Sitter attractors. These results show that boundary-term couplings provide a well-posed, geometrically flexible route to addressing late-time cosmic acceleration.
\end{abstract}

\maketitle

\tableofcontents
\section{Introduction}\label{sec00}

General Relativity (GR) has passed all local and solar-system tests with flying colors, yet on cosmological scales, its vanilla form requires dark components, cold dark matter, and a cosmological constant, to reconcile theory with observations of late-time acceleration and structure growth \cite{Helbig2020,rajantie2012,hawking1966,buchert2016,socas2019}. A particularly systematic pathway to explaining the accelerated expansion of the universe is to alter its matter content by introducing an additional component with a negative pressure, realized through the inclusion of various scalar-field models within the matter sector, such as a canonical scalar field (quintessence) \cite{Ratra1998,Boisseau2000,Guo2007,Dutta2009,Faraoni2000,Gong2002}, a phantom scalar field \cite{Caldwell2002,Kamionkowski2003,Nojiri2003,Saridaki2009}, or a hybrid framework combining both, commonly referred to as quintom models \cite{Piao2005,Zhao2006,Lazkoz2007,Cai2010}. The enormous hierarchy between the observed vacuum energy density and the naive quantum-field theory estimate, together with persistent small tensions in large-scale structure growth and background inferences, motivates exploring well-posed extensions of the gravitational sector itself. A promising approach is the Modified gravity theories that retain diffeomorphism invariance but relax metric compatibility or allow nontrivial geometric scalars beyond the Ricci curvature \cite{bohmer2023,iosifidis2023}. Within the symmetric teleparallel framework where spacetime is flat and torsionless but not metric-compatible, gravity can then be encoded by the non-metricity scalar $Q$, which like the Ricci scalar $\mathring R$ (curvature) and the torsion scalar $\mathbb T$ (teleparallel torsion), yields GR at the level of field equations for a specific Lagrangian \cite{jimenez2018,ad/bianchi}. 
Once a scalar degree of freedom $\phi$ is introduced, the most economical nonminimal couplings are to the geometric scalar itself and to the associated total-divergence (boundary) term, schematically
\begin{align}\label{eqn:ST}
S=\frac{1}{2\kappa }\int\sqrt{-g}\left[f(\phi)Q+\chi(\phi)C-h(\phi)\nabla_\alpha\phi\nabla^\alpha\phi
-U(\phi)+2\kappa\mathcal L_m \right] \,d^{4}x\,,
\end{align}
where $C$ is the boundary term obeying $\mathring R = Q + C$. The pair $\{f(\phi),\chi(\phi)\}$ thus generalizes scalar-tensor gravity to the non-metricity arena\footnote{except when $f=\chi$, in which case the theory reduces to the usual curvature based scalar-tensor gravity}, and the explicit appearance of $C$ is not cosmetic: it alters the geometry through the disformation and superpotential tensors, reshuffling the effective Einstein equations and the scalar dynamics in ways that are essential for the cosmological evolution history. 

There is a close analogue on the metric teleparallel side, where one couples $\phi$ nonminimally to $\mathbb T$ or to the torsional boundary term $B$ entering $\mathring R=-\mathbb T+B$ solely through $\phi^2$ (Jordan-frame) coupling, which helps classify dynamical regimes and late-time attractors \cite{Geng:2011, BahamondeWright:2015}. By contrast, in the present non-metricity-based construction, we are not restricted to power-law choices: generic $f(\phi)$, $\chi(\phi)$, and $U(\phi)$ are admissible, enabling richer phenomenology than the special $\phi^2$ case and allowing the $\{\mathring R,\mathbb T,Q\}$ correspondences to be probed in parallel limits.


Finally, our construction of the non-metricity extension of scalar-tensor gravity with simultaneous couplings to $Q$ and to the boundary term $C$ generalizes earlier $f(Q)$ \cite{Heisenberg2024} and $f(Q,C)$ \cite {De:2023} proposals and their scalar cousins \cite{Jarv:2018,Ghulam2025,murtazasteep}, which are theoretically robust, and observationally well-motivated. The present theory retains GR as a limit, encompasses the teleparallel $B$-coupled scalar as a sister theory in the torsion picture and, thanks to the enlarged functional space for the couplings, offers a flexible yet principled route to model the universe’s expansion and structure formation from a single geometric principle. The explicit boundary-term coupling $\chi(\phi)C$ opens new geometric channels that can mimic early dark energy or realize tracking behavior.

The paper is structured as follows: After the Introduction, we present the mathematical formulation of the symmetric teleparallel framework and derive the corresponding field equations for the non-metricity-boundary coupled scalar-tensor theory in Section~\ref{sec1}. In Section~\ref{CONFORMAL}, we perform the conformal transformation for this theory. In Section~\ref{sec2}, we outline the cosmological implications of this theory within the context of compatible affine connection classes.  Section~\ref{sec3} is devoted to constructing a unified, affine connection-independent autonomous dynamical system in a spatially flat FLRW background. Subsequently, in Subsections~\ref{sub1}, \ref{sub2}, and \ref{sub3}, we respectively perform a comprehensive phase-space analysis for a specific coupling and scalar potential, across the three distinct connection branches. Finally, the main insights of our analysis are summarized in Section~\ref{conlusion}.

\section{The fundamentals} \label{sec1}

In this section, we provide the fundamental mathematical framework for symmetric teleparallel gravity. In the standard geometric description of gravity, the Levi-Civita connection $\mathring{\Gamma}^\alpha{}_{\mu\nu}$ is the unique affine connection characterized by both metric compatibility and vanishing torsion. It is expressed entirely in terms of the metric $g_{\mu\nu}$ as
\begin{equation}
\mathring{\Gamma}^\alpha_{\,\,\,\mu\nu}=\frac{1}{2}g^{\alpha\beta}\left(\partial_\nu g_{\beta\mu}+\partial_\mu g_{\beta\nu}-\partial_\beta g_{\mu\nu} \right)\,.
\end{equation}

In the symmetric teleparallel framework, one considers a more general torsion-free and curvature-free affine connection $\Gamma^\alpha{}_{\mu\nu}$, where the requirement of metric compatibility is relaxed. This leads to a non-vanishing non-metricity tensor defined by
\begin{equation} \label{Q tensor}
Q_{\lambda\mu\nu} = \nabla_\lambda g_{\mu\nu}=\partial_\lambda g_{\mu\nu}-\Gamma^{\beta}_{\,\,\,\lambda\mu}g_{\beta\nu}-\Gamma^{\beta}_{\,\,\,\lambda\nu}g_{\beta\mu}\neq 0 \,.
\end{equation}

The general affine connection can be decomposed into the Levi-Civita part and the disformation tensor $L^\lambda{}_{\mu\nu}$
\begin{equation} \label{connc}
\Gamma^\lambda{}_{\mu\nu} = \mathring{\Gamma}^\lambda{}_{\mu\nu}+L^\lambda{}_{\mu\nu}\,,
\end{equation}
where the disformation tensor is related to the non-metricity tensor via
\begin{equation} \label{L}
L^\lambda{}_{\mu\nu} = \frac{1}{2} (Q^\lambda{}_{\mu\nu} - Q_\mu{}^\lambda{}_\nu - Q_\nu{}^\lambda{}_\mu) \,.
\end{equation}

To construct the gravitational action, we define the non-metricity conjugate (or superpotential) tensor $P^\lambda{}_{\mu\nu}$ as
\begin{equation} \label{P}
P^\lambda{}_{\mu\nu} = \frac{1}{4} \left( -2 L^\lambda{}_{\mu\nu} + Q^\lambda g_{\mu\nu} - \tilde{Q}^\lambda g_{\mu\nu} -\delta^\lambda{}_{(\mu} Q_{\nu)} \right) \,,
\end{equation}
where $Q_\mu$ and $\tilde{Q}_\mu$ represent the two independent traces of the non-metricity tensor
\begin{equation*}
 Q_\mu = g^{\nu\lambda}Q_{\mu\nu\lambda} = Q_\mu{}^\nu{}_\nu \,, \qquad \tilde{Q}_\mu = g^{\nu\lambda}Q_{\nu\mu\lambda} = Q_{\nu\mu}{}^\nu 
\end{equation*}
And subsequently we may define
\begin{align}
 L_\mu = L_\mu{}^\nu{}_\nu \,, \qquad 
 \tilde{L}_\mu = L_{\nu\mu}{}^\nu \,.   
\end{align}
Finally, the non-metricity scalar $Q$, which serves as the basis for the gravitational Lagrangian, is obtained through the contraction
\begin{equation} \label{Q}
Q=Q_{\alpha\beta\gamma}P^{\alpha\beta\gamma}\,.
\end{equation}
Following the curvature-free and torsion-free constraints of $\Gamma$, we can obatin 
\begin{align}
\m R+\m\nabla_\alpha(L^\alpha-\tilde L^\alpha)-Q=0\,. \label{mR}
\end{align}
As $Q^\alpha-\tilde Q^\alpha=L^\alpha-\tilde L^\alpha$, using the last relation (\ref{mR}), we can also express the total divergence (boundary term) as
\begin{align}
C=\m{R}-Q&=-\m\nabla_\alpha(Q^\alpha-\tilde Q^\alpha)
=-\frac1{\sqrt{-g}}\partial_\alpha\left[\sqrt{-g}(Q^\alpha-\tilde Q^\alpha)\right]\,.
\end{align}
The gravitational action integral
$S_{STEGR}=\int\!\!d^{4}x\sqrt{-g}Q$,
leads to a gravitational model equivalent to General Relativity, known as Symmetric Teleparallel Equivalent General Relativity (STEGR). Therefore, an arbitrary function $f(Q)$ of the non-metricity scalar $Q$, was introduced in the action $S_{f(Q)}=\int\!\!d^{4}x\sqrt{-g}f(Q)$, to produce non-trivial gravity models, knows as $f(Q)$ gravity. Note that a linear function $f(Q)$ reduces the theory to $\Lambda$CDM. 
While $f(Q)$ gravity remains a second-order theory analogous to General Relativity (GR), higher-order extensions of the symmetric teleparallel framework can be achieved by incorporating terms such as $\Box^k Q$ into the Lagrangian. However, a more geometrically intuitive approach involves the non-linear inclusion of the boundary term $C$ or its direct coupling with the non-metricity scalar $Q$. This motivation led to the recent proposal of $f(Q,C)$ gravity, characterized by the action
$S=\int d^{4}x\sqrt{-g}\,f(Q,C)$.

Interestingly, our proposed theory with action term (\ref{eqn:ST}) can produce these symmetric teleparallel modified gravity models in certain limits, as shown below. 

\begin{itemize}
\item $\chi\left(  \phi\right)  =\chi_{0}$, the resulting theory is equivalent
to the symmetric teleparallel scalar-tensor theory which includes $f\left(
Q\right)  $ gravity, for $h\left(  \phi\right)  =0$, and $U\left(
\phi\right)  $ nonlinear function.

\item $f(\phi)=f_{0}$ and $h\left(  \phi\right)  =0$, the theory is equivalent
to the separable family of $f\left(  Q,C\right)  =f_{0}Q+F\left(  C\right)  $,
\ theory, with $U\left(  \phi\right)  $ a nonlinear function.

\item $f\left(  \phi\right)  =\chi\left(  \phi\right)  $, the metric
scalar-tensor theory is recovered, which include the $f\left(  R\right)
$ gravity in the case $h\left(  \phi\right)  =0$ and $U\left(  \phi\right)  $
a nonlinear function.
\end{itemize}

The variation of the action term (\ref{eqn:ST}) with respect to the metric produces the metric field equations
\begin{align}\label{eqn:FE1}
\kappa T_{\mu\nu}
=&f\m G_{\mu\nu} +2(f'-\chi')P^\lambda{}_{\mu\nu}\nabla_\lambda \phi
  +(-\m\nabla_\mu\m\nabla_\nu+g_{\mu\nu}\m\nabla^\alpha\m\nabla_\alpha)\chi 
  -h\nabla_\mu\Phi\nabla_\nu\phi
+\frac12hg_{\mu\nu}\nabla^\alpha\phi\nabla_\alpha\phi+\frac12Ug_{\mu\nu} 
\end{align}
In the above expressions, $\mathring{G}_{\mu\nu}$ represents the Einstein tensor associated with the Levi-Civita connection. The energy-momentum tensor $T_{\mu\nu}$ is derived from the matter Lagrangian $\mathcal{L}_M$ through the standard variation
\begin{equation}
T_{\mu\nu}=-\frac{2}{\sqrt{-g}}\frac{\delta(\sqrt{-g}\mathcal{L}_M)}{\delta g^{\mu\nu}} \,, 
\end{equation}
where primes denote differentiation with respect to the scalar field $\phi$. Variation of the action with respect to $\phi$ yields the scalar field equation of motion
\begin{equation} \label{eqn:FE2}
f'Q+\chi'C+h'\nabla^\alpha\phi\nabla_\alpha\phi+2h\mathring{\nabla}^\alpha\mathring{\nabla}_\alpha \phi-U'=0 \,. 
\end{equation}
In addition to the metric and the scalar field, the components of the affine connection constitute an independent set of dynamical variables. Their variation leads to the connection field equations
\begin{align}\label{eqn:FE3}
(\nabla_\mu-\tilde L_\mu)(\nabla_\nu-\tilde L_\nu)
\left[4(f-\chi)P^{\mu\nu }{}_\lambda+\kappa\Delta_\lambda{}^{\mu\nu}\right]=0\,,
\end{align}
where 
\[\Delta_\lambda{}^{\mu\nu}=-\frac2{\sqrt{-g}}\frac{\delta(\sqrt{-g}\mathcal L_M)}{\delta\Gamma^\lambda{}_{\mu\nu}}\,,\]is the hypermomentum tensor.


Unlike the scalar-tensor extension of GR, the non-metricity approach is not straightaway compatible with the classical energy conservation condition. 
The divergence of  (\ref{eqn:FE1}) yields  
\begin{align}\label{vv0}
\kappa\m\nabla_\mu T^\mu{}_\nu
=&\m G^\lambda{}_\nu\nabla_\lambda (f-\chi)
+2\m\nabla_\mu\left(P^{\lambda\mu}{}_\nu\nabla_\lambda (f-\chi)\right)
-\frac12\m R\nabla_\nu \chi
 +\left(-h\m\nabla^\alpha\nabla_\alpha\phi 
        -\frac12h'\m\nabla^\alpha\m\nabla_\alpha\phi+\frac12U'\right)\phi_\nu\,.
\end{align}
In view of (\ref{eqn:FE2}), the equation (\ref{vv0}) becomes 
\begin{align}
\kappa\m\nabla_\mu T^\mu{}_\nu
 =&\left(\m G^\lambda{}_\nu+\frac Q2\delta^\lambda{}_\nu\right) \nabla_\lambda (f-\chi)
 +2\m\nabla_\mu\left(P^{\lambda\mu}{}_\nu\nabla_\lambda (f-\chi)\right)\,.
\end{align}
We utilise the following result
\begin{align}
2(\nabla_\lambda&-\tilde L_\lambda)(\nabla_\mu-\tilde L_\mu)
 \left((f-\chi) P^{\lambda\mu}{}_\nu\right) 
 =\left(\m G^\lambda{}_\nu+\frac Q2\delta^\lambda{}_\nu\right) \nabla_\lambda (f-\chi)
 +2\m\nabla_\mu\left(P^{\lambda\mu}{}_\nu\nabla_\lambda (f-\chi)\right)\,,
 \label{eqn:C1a}
\end{align}
to finally conclude that

\begin{align}\label{eqn:EC}
\kappa\m\nabla_\mu T^\mu{}_\nu
=&2(\nabla_\lambda-\tilde L_\lambda)(\nabla_\mu-\tilde L_\mu)
 \left((f-\chi) P^{\lambda\mu}{}_\nu\right)=0
\end{align}
in view of (\ref{eqn:FE3}), if either the hypermomentum tensor is conserved or the matter Lagrangian is independent of the affine connection \cite{Hohmann:gencov}.


\section{Conformal Transformation}\label{CONFORMAL}
Conformal transformations, defined by the rescaling, preserve the causal structure of spacetime and therefore provide a powerful geometric tool in gravitational physics and cosmology. Their importance is particularly evident in scalar-tensor theories \cite{Faraoni2004,Kaiser2010}, where a nonminimal coupling between the scalar field and the metric complicates the gravitational sector. By applying a conformal transformation, this coupling can be removed, mapping the theory into the Einstein frame, where gravity appears in its standard form and the scalar field becomes minimally coupled. Although the transformation ensures a one-to-one correspondence between solution trajectories of the field equations in both frames, physical quantities constructed from those solutions are not generally invariant \cite{Gionti:2017ffe,Dimakis:2023cam}. As a result, singular solutions in one frame may correspond to nonsingular ones in the other \cite{Gunzig1999,SJ2021}, and observable quantities such as masses, distances, or scalar potentials may differ. These subtleties explain why debates regarding a “preferred” frame continue and why Hamiltonian formulations of the two frames can be inequivalent \cite{Gionti2024}. Nonetheless, conformal mappings remain essential for generating new exact solutions, simplifying cosmological dynamics, and analysing the evolution of scalar-curvature and scalar non-metricity theories, where they often preserve the major cosmological epochs despite altering detailed physical interpretations. Let us provide a conformal transformation for our present theory.

Let $\bar{g}_{\mu\nu}$, $g_{\mu \nu}$ be two metric tensors that share the same conformal algebra, meaning that the metrics are conformally related in such a way that
\begin{align}
    \bar{g}_{\mu\nu}=e^{2\Omega(x^k)}g_{\mu \nu}, ~~ \bar{g}^{\mu\nu}=e^{-2\Omega(x^k)}g^{\mu \nu}
\end{align}
where $\Omega(x^k)$ is the so-called conformal function.

The nonmetricity tensors $\bar{Q}_{\lambda\mu\nu}$ and ${Q}_{\lambda\mu\nu}$, corresponding to
the two conformally related metrics, are related as
\begin{equation}
\bar{Q}_{\lambda\mu\nu}=e^{2\Omega}{Q}_{\lambda\mu\nu}+2\Omega_{,\lambda}%
\bar{g}_{\mu\nu}%
\end{equation}

and the respective nonmetricity scalars as
\begin{equation}
\bar{Q}=e^{-2\Omega}Q+(2\Omega_{,\lambda}P^{\lambda}+6\Omega_{,\lambda}%
\Omega^{,\lambda})
\end{equation}

Hence, the corresponding boundary terms are related as follows%
\begin{equation}
\bar{C}=e^{-2\Omega}C+6g^{\mu\nu}\mathring{\nabla}_{\mu}%
\mathring{\nabla}_{\nu}\Omega-2P^{\lambda}\Omega
_{,\lambda}-12\,\Omega_{,\lambda}\Omega^{,\lambda}%
\end{equation}
Consider now the Action Integral (\ref{eqn:ST}) for the metric tensor $\bar
{g}_{\mu\nu}$, that is,
\begin{equation}
\bar{S}=\int\sqrt{-\bar{g}}\left[  \frac{f(\phi)}{2}\bar{Q}+\frac{\chi(\phi
)}{2}\bar{C}-\frac{h(\phi)}{2}\bar{g}^{\mu\nu}\phi_{,\mu}\phi_{,\nu}%
-\frac{U(\phi)}{2}\right]  \,d^{4}x\,.
\end{equation}

Therefore, the latter Action Integral in terms of the scalars for the
conformally related metric $g_{\mu\nu}$, with $\Omega=-\frac{1}{2}\ln f\left(
\phi\right)  $, reads%
\begin{equation}
\bar{S}\int\sqrt{-\bar{g}}\left[  \frac{Q}{2}+A\left(  \phi\right)  \bar
{C}-\frac{1}{2}H\left(  \phi\right)  \bar{g}^{\mu\nu}\phi_{,\mu}\phi_{,\nu
}-V\left(  \phi\right)  \right]  ,
\end{equation}
where~%
\begin{equation}
V\left(  \phi\right)  =\frac{U\left(  \phi\right)  }{2f^{2}\left(
\phi\right)  },
\end{equation}
\begin{equation}
A\left(  \phi\right)  =\left(  \frac{\chi(\phi)}{f(\phi)}+(f(\phi)-\chi
(\phi))\frac{\ln f(\phi)}{2f^{2}(\phi)}\right)  ,
\end{equation}
and%
\begin{equation}
H\left(  \phi\right)  =\frac{1}{f\left(  \phi\right)  }\left(  \frac
{3f_{,\phi}^{2}}{2f\left(  \phi\right)}-h(\phi)\right)  +3\frac{\chi(\phi)}{f^{2}\left(
\phi\right)  }\left(  f_{,\phi\phi}+f_{,\phi}-2\frac{f_{,\phi}^{2}}{f\left(
\phi\right)  }\right)  .
\end{equation}

Thus, after the conformal transformation in (\ref{eqn:ST}) we can recover the
following theories

\begin{itemize}
\item $A\left(  \phi\right)  =0$, STEGR/GR is recovered with a scalar field.

\item $H\left(  \phi\right)  =0$, $f\left(  Q,C\right)  $ theory of gravity is
recovered, for $V\left(  \phi\right)  $ a nonlinear function.
\end{itemize}

We remark that after a second conformal transformation, it is possible to
recover $f\left(  Q\right)  $ gravity, or other non-metricity scalar-tensor theories.

It is well known, that the conformal transformation in non-metricity
scalar-tensor theory does not lead to STEGR, unlike the case of metric
scalar-tensor theory that recover GR. However, when the boundary term is
introduced in the non-metricity scalar-tensor theory, the conformal
transformation leads to a theory in the Einstein frame that can describe STEGR.

\section{Cosmological application}\label{sec2}
We consider a homogeneous and isotropic Friedmann-Lema\^{i}tre-Robertson-Walker (FLRW) spacetime given by 
\begin{align}\label{ds:RW}
ds^2=-dt^2+a^2\left(\frac{dr^2}{1-kr^2}+r^2d\theta^2+r^2\sin^2\theta d\phi^2\right)
\end{align}
with scale factor $a(t)$, Hubble parameter $H=\dot a/a$ and the spatial curvature $k=0,+1,-1$ respectively 
modeled the universe of spatially flat, closed, and open types. In the present work, we study a spatially flat spacetime.
Here the $\dot{(~)}$ denotes the derivative with respect to $t$.
In addition, we denote $u_\mu=(dt)_\mu$ and  
$h_{\mu\nu}=g_{\mu\nu}+u_\mu u_\nu$. We consider a perfect fluid type stress energy tensor given by
\begin{align}
T_{\mu\nu}=pg_{\mu\nu}+(p+\rho)u_\mu u_\nu
\end{align}
where $\rho$ and $p$  denote 
the energy density and the pressure, respectively.

There are three classes of affine connections that are compatible with the symmetric teleparallel framework, which are given as follows \cite{FLRW/connection}
\begin{align} \label{eqn:conn}
\Gamma^t{}_{tt}=&C_1, 
	\quad 					\Gamma^t{}_{rr}=\frac{C_2}{1-kr^2}, 
	\quad 					\Gamma^t{}_{\theta\theta}=C_2r^2, 
	\quad						\Gamma^t{}_{\phi\phi}=C_2r^2\sin^2\theta,								\notag\\
\Gamma^r{}_{tr}=&C_3, 
	\quad  	\Gamma^r{}_{rr}=\frac{kr}{1-kr^2}, 
	\quad		\Gamma^r{}_{\theta\theta}=-(1-kr^2)r, 
	\quad		\Gamma^r{}_{\phi\phi}=-(1-kr^2)r\sin^2\theta,												\notag\\
\Gamma^\theta{}_{t\theta}=&C_3, 
	\quad		\Gamma^\theta{}_{r\theta}=\frac1r,
	\quad		\Gamma^\theta{}_{\phi\phi}=-\cos\theta\sin\theta,										\notag\\
\Gamma^\phi{}_{t\phi}=&C_3, 
	\quad 	\Gamma^\phi{}_{r\phi}=\frac1r, 
	\quad 	\Gamma^\phi{}_{\theta\phi}=\cot\theta,
\end{align}
where  $C_1$, $C_2$ and $C_3$ are functions of $t$, which must fulfill the following criteria\footnote{Only 2 of these 3 equations are independent as eq (\ref{c2c3}) can be straightaway identified as a derivable from eq (\ref{dotc3}) and (\ref{dotc32}).}  
\begin{align}\label{c2c3}
C_2C_3+k=&0\,,
\end{align}
\begin{align}\label{dotc3}
\dot C_2+C_2(C_1-C_3)=&0\,,
\end{align}
\begin{align}\label{dotc32}
\dot C_3+C_3(C_3-C_1)=&0\,.
\end{align}
The above geometrical constraints (\ref{c2c3})-(\ref{dotc32}) play a pivotal role in the framework of the autonomous system for the generic class in a unified manner, as we can see in the next section. 
\begin{enumerate}\label{casesConn}
\item[(I)] $C_1=\gamma$, $C_2=C_3=0$ and $k=0$, where $\gamma$ is a temporal function; or
\item[(II)] $C_1=\gamma+\dfrac{\dot\gamma}\gamma$, $C_2=0$, $C_3=\gamma$ and $k=0$,
             where $\gamma$ is a nonvanishing temporal function; or 
\item[(III)] $C_1=-\dfrac k\gamma-\dfrac{\dot\gamma}{\gamma}$, $C_2=\gamma$, $C_3=-\dfrac k\gamma$ and $k=0,\pm1$,
             where $\gamma$ is a nonvanishing temporal function.
\end{enumerate} 
Let us denote these 3 connection classes by $\Gamma_A$, $\Gamma_B$ and $\Gamma_C$. Hence within the formulation (\ref{eqn:conn}),
the disformation tensor; the superpotential tensor, the non-metricity scalar and the boundary term can be derived respectively as 
\begin{align}
L^\lambda{}_{\mu\nu}
=&C_1u^\lambda u_\mu u_\nu
+\left(\frac{C_2}{a^2}-H\right)u^\lambda h_{\mu\nu}
+(C_3-H)(-h^\lambda{}_\nu u_\mu-h^\lambda{}_\mu u_\nu)  \label{eqn:L2}\\
2P^\lambda{}_{\mu\nu}
=&\frac12\left(3C_3-3\frac{C_2}{a^2}\right)u^\lambda u_\mu u_\nu
    +\frac12\left(3C_3+\frac{C_2}{a^2}-4H\right)u^\lambda h_{\mu\nu}
    +\frac12(C_1+C_3-H)(-h^\lambda{}_\mu u_\nu-h^\lambda{}_\nu u_\mu) 
    \label{eqn:P2}\\
Q
=&3\left(-2H^2+2\frac k{a^2}+3HC_3+\dot C_3+\frac1{a^2}(HC_2+\dot C_2)\right)\,,
\label{eqn:Q2}\\
C
=&3\left(2\dot{H}+6H^2-2\frac k{a^2}-3HC_3-\dot C_3-\frac1{a^2}(HC_2+\dot C_2)\right)\,.
\end{align}
The Friedmann-like equations generically can be derived\footnote{The specific equations of motion for each connection class can be found in Appendices \ref{GammaA}, \ref{GammaB} and \ref{GammaC}, respectively.}
\begin{align}\label{p_general}
\kappa p
=f\left(-2\dot H-3H^2-\frac k{a^2}\right)
    +\frac12(\dot f-\dot \chi) \left(3C_3+\frac{C_2}{a^2}-4H\right)
    -\frac12h\dot \phi^2    +\frac12U -2H\dot \chi-\ddot\chi \,,
\end{align}
\begin{align}\label{rho_general}
\kappa\rho
=f\left(3H^2+3\dfrac k{a^2}\right)
    +\frac12(\dot f-\dot\chi) \left(3C_3-3\frac{C_2}{a^2}\right)
     -\frac12h\dot \phi^2 -\frac12U+3H\dot \chi \,,
\end{align}

The scalar field equation (\ref{eqn:FE2}) provides
\begin{align}\label{eqn:FE2cosmo}
0=&3f'\left(-2H^2+2\frac k{a^2}+3HC_3+\dot C_3+\frac1{a^2}(HC_2+\dot C_2)\right)\notag\\
&+3\chi' \left(2\dot{H}+6H^2-2\frac k{a^2}-3HC_3-\dot C_3-\frac1{a^2}(HC_2+\dot C_2)\right) \notag\\
&+h'\nabla^\alpha\phi\nabla_\alpha\phi+2h\m\nabla^\alpha\m\nabla_\alpha \phi-U'.
\end{align}

Moreover, the connection field equation can be directly deduced from (\ref{eqn:FE3})

\begin{align}\label{connfe}
\frac32\left[
    (\dot f-\dot\chi)\left(3C_3H-\frac{2\dot C_2+C_2H}{a^2}\right)
    +(\ddot f-\ddot\chi)\left(C_3-\frac{C_2}{a^2}\right)\right]=0.
\end{align}
Combining, we obtain the usual continuity relation
\begin{align}
    \kappa\dot{\rho}+3\kappa H(\rho+p)=0.
\end{align}

\section{A connection-independent unified autonomous system}\label{sec3}
The dynamical systems approach plays a central role in modified gravity theories \cite{Boehmer:2023,Hussain2024jcap,Pradhan2025,Billyard:2000bh,Coley:1999mj,Paliathanasis2014,Leon2009,Paliathanasis:2024abl,muratazacurvature,vanderWesthuizen:2025vcb,Caldera-Cabral:2010yte,Carloni:2024ybx,murtazafQC,Duchaniya:2024vvc,Murtaza:2025gme,Shabani:2025qxn}. Casting the cosmological equations into an autonomous system of dimensionless variables allows a global phase-space view: critical points (CPs) encode exact or asymptotic cosmological solutions (radiation, matter, scaling, accelerating), their eigenvalues diagnose stability and track viable thermal histories (radiation $\rightarrow$ matter $\rightarrow$ acceleration). This birds-eye analysis is particularly valuable when the theory space includes multiple couplings $\{f,\chi,h,U\}$, nontrivial affine-connection branches, and boundary contributions, since it cleanly separates universal features (existence and stability of attractors) from model-specific details.

In this section, we prescribe a unified framework that yields a connection-independent autonomous dynamical system for the underlying gravity theory, valid for general potentials and coupling functions. Concrete phase-space analyses corresponding to different connection classes are subsequently presented as case studies using a simple ansatz, assuming a spatially flat ($k=0$) FLRW universe, in which case the Friedmann-like equations (\ref{p_general}) and (\ref{rho_general}) reduce to
\begin{align}\label{peq}
\kappa p
=&f\left(-2\dot H-3H^2\right)
    +\frac12(\dot f-\dot \chi) \left(3C_3+\frac{C_2}{a^2}-4H\right)
    -\frac12h\dot \phi^2    +\frac12U -2H\dot \chi-\ddot\chi \,,\\
\kappa\rho
=&f\left(3H^2\right)
    +\frac12(\dot f-\dot\chi) \left(3C_3-3\frac{C_2}{a^2}\right)
     -\frac12h\dot \phi^2 -\frac12U+3H\dot \chi \,,
\end{align}
The scalar field eq for this setup from Eq (\ref{eqn:FE2cosmo}) can be written ($h$ a constant)
\begin{align}\label{sclreq}
f'\left(-6H^2+9HC_3+3\dot C_3+\frac{3HC_2}{a^2}+\frac{3\dot{C_2}}{a^2}\right)+\chi'\left(6\dot{H}+18H^2-9HC_3-3\dot C_3-\frac{3HC_2}{a^2}-\frac{3\dot{C_2}}{a^2}\right)-2h(\ddot\phi+3H\dot\phi)-U'=0.
\end{align}
In the context of $H$-normalization, we consider the dimensionless variables,
\begin{align}
\label{variables}
  x_1 &= \frac{\kappa \rho }{3H^2f},\quad
  x_2 = \frac{\dot{\phi}}{\sqrt{6f}H},\quad
    x_3 = \frac{U}{3H^2f}, \quad
   x_4 = \frac{C_2}{a^2H},\quad
    x_5 = \frac{C_3}{H},\quad
     \mu = \frac{\chi'}{\sqrt{f}},\notag \\
  \lambda& = \frac{U' \sqrt{f}}{U}, \quad
    \zeta = \frac{f'}{\sqrt{f}}, \quad
     \eta = \frac{\chi'}{\sqrt{\chi}},\quad
   \Gamma  = \frac{U'' U}{U'^2}, \quad
   \Delta = \frac{f''f}{f'^2}, \quad
   \Theta=\frac{\chi \chi''}{\chi'^2}.
\end{align}


The constraint equation can be provided as
\begin{align}\label{generic_constraint}
    x_1=1 + \sqrt{6}\,\mu\,x_2 - h\,x_2^2 - \frac{x_3}{2}+ \sqrt{\frac{3}{2}}\,\mu\,x_2\,x_4- \sqrt{\frac{3}{2}}\,\zeta\,x_2\,x_4- \sqrt{\frac{3}{2}}\,\mu\,x_2\,x_5+ \sqrt{\frac{3}{2}}\,\zeta\,x_2\,x_5.
\end{align}

After some computations and by using Eq (\ref{dotc3}), (\ref{dotc32}), (\ref{connfe}), (\ref{peq}), and (\ref{sclreq}) as required, a unified autonomous dynamical system for a generic affine connection
class can be given in terms of $E_i(x_1,x_2,x_3,x_4,x_5,\mu,\lambda,\zeta,\eta;h)$, see Appendix \ref{autom} for details,
\begin{align}
\frac{dx_i}{dN}=E_i,\quad
\frac{d\mu}{dN}=E_6,\quad 
\frac{d\lambda}{dN}=E_7,\quad
\frac{d\zeta}{dN}=E_8,\quad
\frac{d\eta}{dN}=E_9.
\end{align}
We can also introduce the deceleration parameter $q=-1-\frac{\dot{H}}{H^2}$\footnote{The expression can be found in Appendix \ref{autom},} 
to determine whether the universe is undergoing an acceleration phase.

Finally, to exhibit the phase portraits of a particular coupling and potential ansatz, we choose $\mu, \zeta, \lambda$ and $\eta $ to be constants\footnote{By choosing a constant $\mu$ or $\eta$ provide an almost similar form for $\chi$. Note that, our primary aim in this study is to introduce the new nonmetricity-boundary-scalar field coupled gravity theory, its cosmological implication, and finally a prescription of how one can construct an autonomous dynamical system in a unified model-independent, connection-independent way in terms of the standard dimensionless dynamical variables. Thereafter to proceed with the phasespace analysis once the couplings and potential are provided. So the present ansatz is considered as the simplest example, and exponential law or more complicated forms can be analyzed in future.} and this gives the following form for $f, U$ and $\chi$. 
\begin{align}\label{assumption}
    f(\phi)=\frac{\zeta_0^2\phi^2}{4},~U(\phi)=U_0\phi^{\frac{2\lambda_0}{\zeta_0}},~\chi=\frac{\mu_0\zeta_0 \phi^2}{4}
\end{align}
Without loss of generality, we set $\zeta_0,\eta_0=1$. Under the above assumptions and using the constraint equation, the unified four-dimensional autonomous dynamical system can be expressed as
\begin{align}
\frac{dx_i}{dN}=\mathbb{E}_i(x_2,x_3,x_4,x_5;\mu_0,\lambda_0,h),\quad i=2,3,4,5.
\end{align}
For a smooth flow of the text, we use notations $E_i$ and $\mathbb{E}_i$; the detailed expressions of the autonomous system are provided in the Appendix \ref{autom}.

\subsection{Phasespace and stability analysis of $\Gamma_A$}\label{sub1}
For the connection $\Gamma_A$, under the assumptions (\ref{assumption}), and using the constraint equation (\ref{generic_constraint}), we can get the two-dimensional autonomous dynamical system from equations (\ref{generic_eq1}) and (\ref{generic_eq4}) by adopting the following method: firstly put $x_5=0$, then apply L'Hospital's rule, and then substitute $x_4=0$ in the previously mentioned equations. The four critical points obtained for the two-dimensional dynamical system are summarized in Table \ref{table1_conl} and Table \ref{table2_conl}.

The critical point $P_1$ corresponds to a solution in which the dynamics of the scalar field are influenced by the parameters $\mu_0$ and $h$. The deceleration parameter at this point is given by $q=\frac{-2+h+4\mu_0}{2(h+\mu_0)}$. This expression shows that the expansion rate of the universe at $P_1$ depends sensitively on both $\mu_0$ and $h$. In particular, $P_1$ represents an accelerated phase of cosmic expansion when the parameters satisfy the condition $(h<-\frac{2}{3} \land\frac{2-h}{4}<\mu_0<-h)$. Furthermore, it exhibits stable behavior when ($\lambda_0<-1\land h<-\frac{2}{3} \land\mu_0>\frac{2}{3}$).

The second critical point $P_2$ represents a scaling solution, exhibiting stable behavior when $(\lambda_0 >1 \land~ h=-\frac{2}{3} \land \mu_0<\frac{4}{3(1+\lambda_0)})$. The parametric dependent deceleration parameter corresponding to this point is given by $q=\frac{-1-2h+\lambda_0^2+2\mu_0-4\lambda_0\mu_0}{2h+\mu_0+\lambda_0\mu_0}$. Moreover, $P_2$ can also describe an accelerating universe when the parameters satisfy $(\mu_0=1 \land h=-1 \land \lambda_0<1 \lor 1<\lambda_0<3 )$.

The coordinates of the critical point $P_{+3}$ are $\left(\frac{\sqrt{3}\mu_0 +\sqrt{2h+3\mu_0^2}}{\sqrt{2}h},0\right)$, this point gives a solution where the dynamics of the scalar field are governed by the parameters $\mu_0$ and $h$. This point exists for non-vanishing values of $h \neq 0$. The deceleration parameter $q$ at point is expressed as $q=\frac{2h+3\mu_0+\sqrt{6h+9\mu_0^2}}{h}$. The point $P_{3+}$ accounts for an accelerated phase of the universe when $(\mu_0=1\land -\frac{3}{2}<h<0)$ and exhibits stability under the constraints $(1<\lambda_0<3\land 0<\mu_0<\frac{2}{3}\land -\frac{3\mu_0^2}{2}\leq h<0)$.

This point $P_{3-}$ also yields a solution in which the dynamics of the scalar field are primarily driven by the parameters $\mu_0$ and $h$. For this point, $q$ can be obtained as  $q=\frac{2h+3\mu_0-\sqrt{6h+9\mu_0^2}}{h}$. This point also exists for non-zero $h$ and leads to cosmic acceleration when $(\mu_0=\frac{1}{2}\land -\frac{3}{8}\leq h<0)$. This point remains stable for parameter values satisfying ($\mu_0<\frac{1}{3} \land \lambda_0>1 \land h>0)$.

\begin{figure}[tbp]
    \centering
    \begin{subfigure}[b]{0.45\textwidth}
        \centering
        \includegraphics[width=\textwidth]{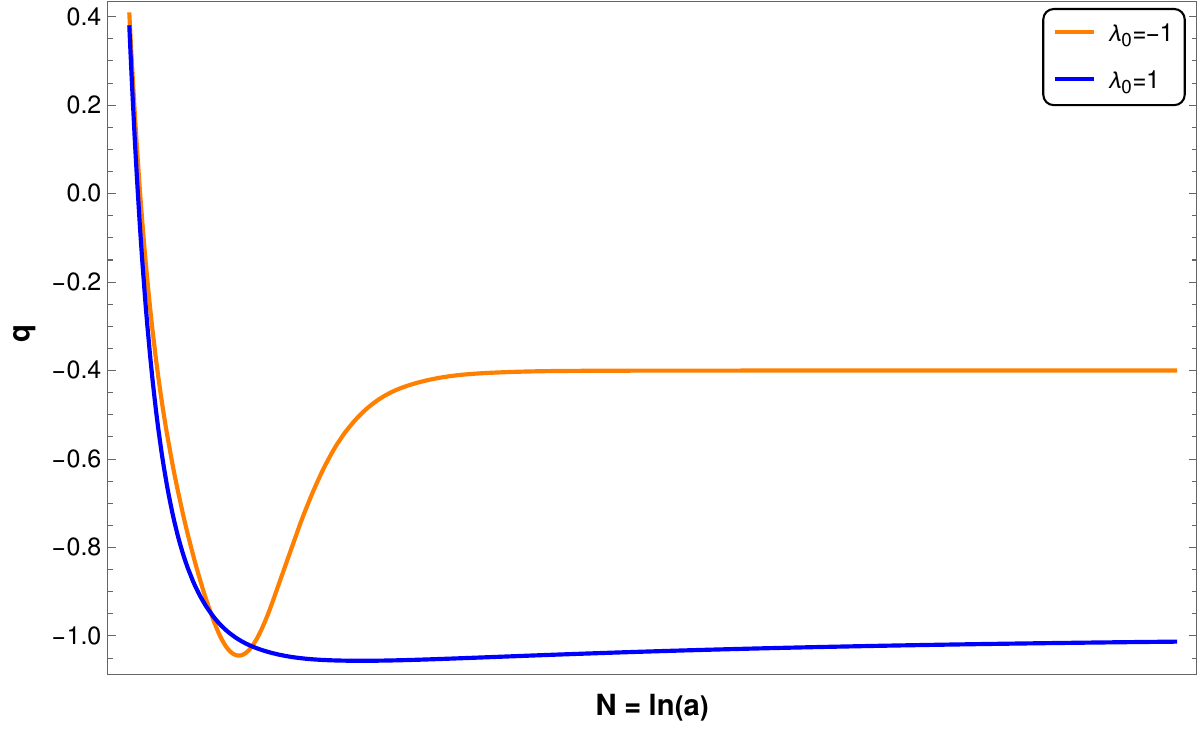}
    \end{subfigure}
    \begin{subfigure}[b]{0.45\textwidth}
        \centering
        \includegraphics[width=\textwidth]{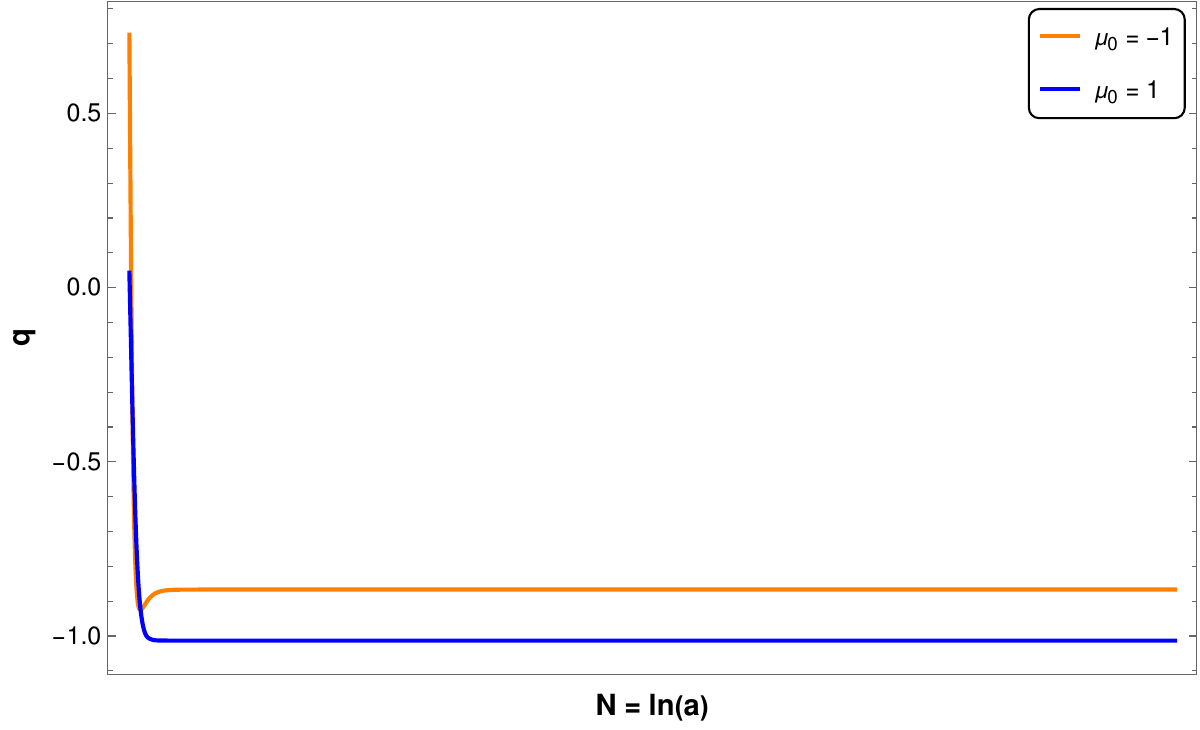}
    \end{subfigure}
    
    \caption{Qualitative evolution of the deceleration parameter for connection $\Gamma_A$ for different values of $\lambda_0$ and $\mu_0$, with initial conditions ($x_2[0]=0.1,~x_3[0]=0.3$) with parameter value is  $h=0.5$.}
    \label{fig1_conl}
\end{figure}

The qualitative evolution of the deceleration parameter of the dynamical system for connection $\Gamma_A$ for different values of $\lambda_0$ and $\mu_0$ is plotted in Fig \ref{fig1_conl}.

\begin{table}[h]
    \centering
      \begin{tabular}{|c|c|c|c|}
        \hline
        Critical point  &$(x_2,x_3)$ &Existence & $q$ \\ \hline
      $P_1$ & $\left(\frac{-2+3\mu_0}{\sqrt{6}(h+\mu_0)},0\right)$ & $h+\mu_0 \neq 0$&$\frac{-2+h+4\mu_0}{2(h+\mu_0)}$\\ \hline
             $P_2$ & $\left(\frac{\sqrt{\frac{2}{3}}(3\mu_0-1-\lambda_0)}{2h+\mu_0+\lambda_0 \mu_0},\frac{2(6h-(1+\lambda_0)(1+\lambda_0-6\mu_0))(2h+3\mu_0^2)}{3(2h+\mu_0+\lambda_0\mu_0)^2}\right)$ & $2h+\mu_0+\lambda_0\mu_0 \neq 0$&$\frac{-1-2h+\lambda_0^2+2\mu_0-4\lambda_0\mu_0}{2h+\mu_0+\lambda_0\mu_0}$\\ \hline
              $P_{\pm3}$ & $\left(\frac{\sqrt{3}\mu_0 \pm\sqrt{2h+3\mu_0^2}}{\sqrt{2}h},0\right)$ &$h\neq 0$&$\frac{2h+3\mu_0\pm\sqrt{6h+9\mu_0^2}}{h}$\\ \hline
          \end{tabular}
    \caption{Critical points and their physical properties.}
    \label{table1_conl}
\end{table}

\begin{table}[H]
\centering
\scriptsize
\begin{tabular}{|c|c|c|}
\hline
 Critical point  & Eigenvalues & Stability \\ \hline
$P_{1}$ & $\left( -3+\frac{2+3h}{2(h+\mu_0)}, \frac{3h-2\lambda_0+3(1+\lambda_0)\mu_0}{(h+\mu_0)}\right)$ & stable for $\lambda_0<-1\land h<-\frac{2}{3} \land\mu_0>\frac{2}{3}$ \\ \hline
 $P_{2}$ & $\left(  -6+\frac{6h+(1+\lambda_0)^2}{(2h+\mu_0+\lambda_0\mu_0)},\frac{-6h+2\lambda_0(1+\lambda_0)-3(1+3\lambda_0)\mu_0}{(2h+\mu_0+\lambda_0\mu_0)}\right)$ &stable for $\lambda_0 >1 \land~ h=-\frac{2}{3} \land \mu_0<\frac{4}{3(1+\lambda_0)}$ \\ \hline
        $P_{\pm3}$ & $\left(\frac{3h+3\mu_0\pm\sqrt{6h+9\mu_0^2}}{h},6\pm\frac{(1+\lambda_0)(\pm3\mu_0+\sqrt{6h+9\mu_0^2})}{h}\right)$& stable for $[1<\lambda_0<3\land 0<\mu_0<\frac{2}{3}\land -\frac{3\mu_0^2}{2}\leq h<0]_{+}$, stable for $[\mu_0<\frac{1}{3} \land \lambda_0>1 \land h>0]_{-}$\\\hline
    \end{tabular}
    \caption{Eigenvalues and Stability.}
    \label{table2_conl}
\end{table}

\subsection{Phasespace and stability analysis of $\Gamma_B$}\label{sub2}
To construct the dynamical system for this connection branch $\Gamma_B$, we take $x_4=0$. Under the assumptions (\ref{assumption}) and constraint equation (\ref{generic_constraint}), we get three-dimensional dynamical system from equations (\ref{generic_eq1}) and (\ref{generic_eq4}). The details of the critical points (CP) for three-dimensional dynamical system are provided in Table \ref{table1_conll} and Table \ref{table2_conll}, including their existence conditions, deceleration parameter $q$, and stability conditions.  

The critical point $P_1$ exists for $\mu_0 \neq1$. The corresponding deceleration parameter is $q=\frac{1}{2}$, which characterizes a decelerating, matter (dust)-dominated epoch of the universe. This is consistent with the standard cosmological model during the matter-dominated era, where pressureless matter governs the cosmic dynamics, and the expansion slows down due to gravitational attraction. The eigenvalues of the linearized dynamical system around $P_1$ are $\left(-\tfrac{3}{2},-\tfrac{3}{2}, 3 \right)$, indicating that the point is a saddle point and suggesting that the universe can temporarily evolve near this matter-dominated state before moving toward a late-time attractor corresponding to accelerated expansion.

The second critical point, $P_2\left( -\frac{\sqrt{6}}{1+\lambda_0},0,2+\frac{6h-(\lambda_0-5)(\lambda_0+1)}{3(\lambda_0+1)(\mu_0-1)} \right)$, exists under the conditions ( $\lambda_0 \neq -1 \land \mu_0 \neq 1$). The deceleration parameter at this point is given by $q=2-\frac{6}{1+\lambda_0}$. Acceleration occurs when $(-1<\lambda_0<2)$, corresponding to $q<0$. Therefore, this critical point can describe an accelerated phase of cosmic expansion, potentially associated with an inflationary or dark energy-dominated regime, depending on parameter values. The stability conditions for the point $P_2$ are provided in Table \ref{table2_conll}. 

The third critical point, $P_3\left(0,2,2-\frac{2(\lambda_0-2)}{3(\mu_0-1)} \right)$), exists for $\mu_0 \neq1$ and represents a late-time de Sitter attractor characterized by $q=-1$. This value of the deceleration parameter corresponds to an exponentially expanding universe, similar to the observed present-day acceleration attributed to dark energy. The eigenvalues of the linearized system around $P_3$ are all negative, $\left( -3,-3,-3 \right)$, confirming that $P_3$ is a stable node or attractor. Hence, the universe naturally evolves toward this point in the asymptotic future, making it a strong candidate for the dark energy-dominated era.

The critical point $P_4\left(x_2,0,\frac{\sqrt{\frac{2}{3}}(1-hx_2^2+\sqrt{6}x_2\mu_0)}{x_2(\mu_0-1)}\right)$ represents a one parameter family of solutions that exists for $(x_2 \neq 0\land\mu_0 \neq 1)$. The cosmological characteristics of this point depend explicitly on the value of $x_2$. In particular, the point can account for cosmic acceleration when 
$(x_2<-\sqrt{\frac{2}{3}})$. Such behavior suggests that this family may encompass a wide range of cosmological scenarios, from decelerating to accelerating regimes, depending on the interplay between the parameters $h,\mu_0$, and $x_2$. Therefore, $P_4$ can be interpreted as a generalized scaling solution, potentially interpolating between matter-dominated and accelerated phases.

\begin{figure}[tbp]
    \centering
    \begin{subfigure}[b]{0.45\textwidth}
        \centering
        \includegraphics[width=\textwidth]{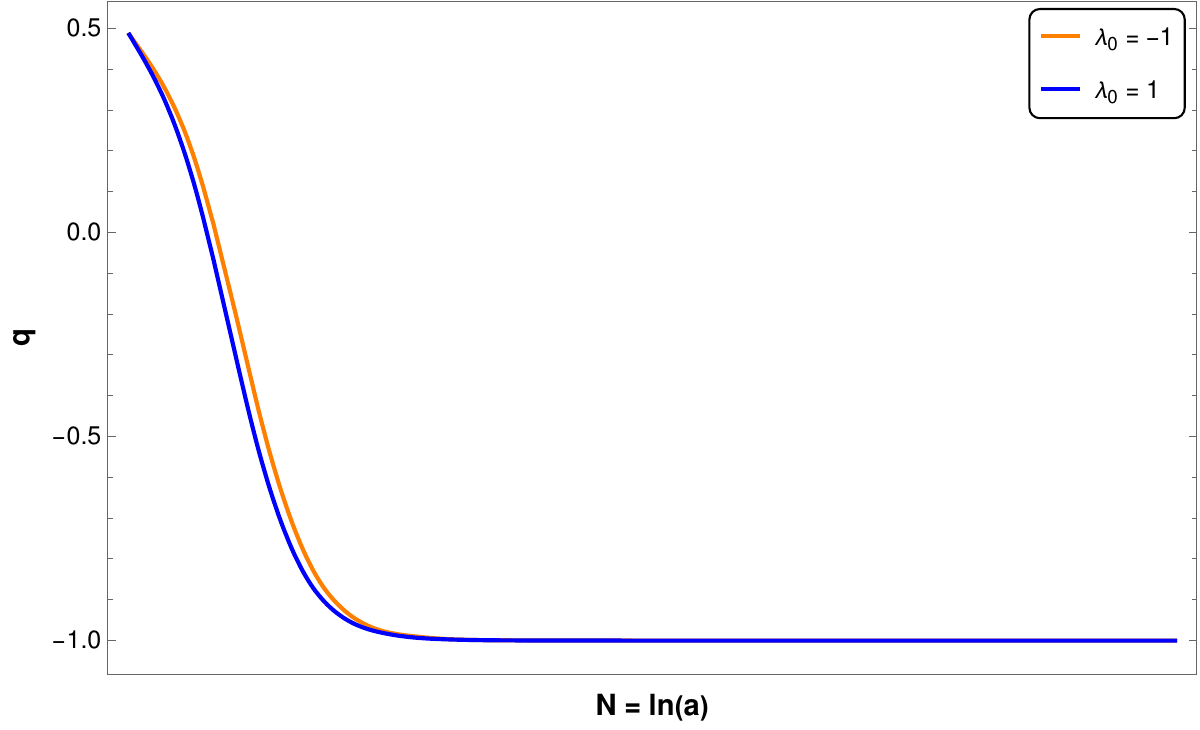}
    \end{subfigure}
    \begin{subfigure}[b]{0.45\textwidth}
        \centering
        \includegraphics[width=\textwidth]{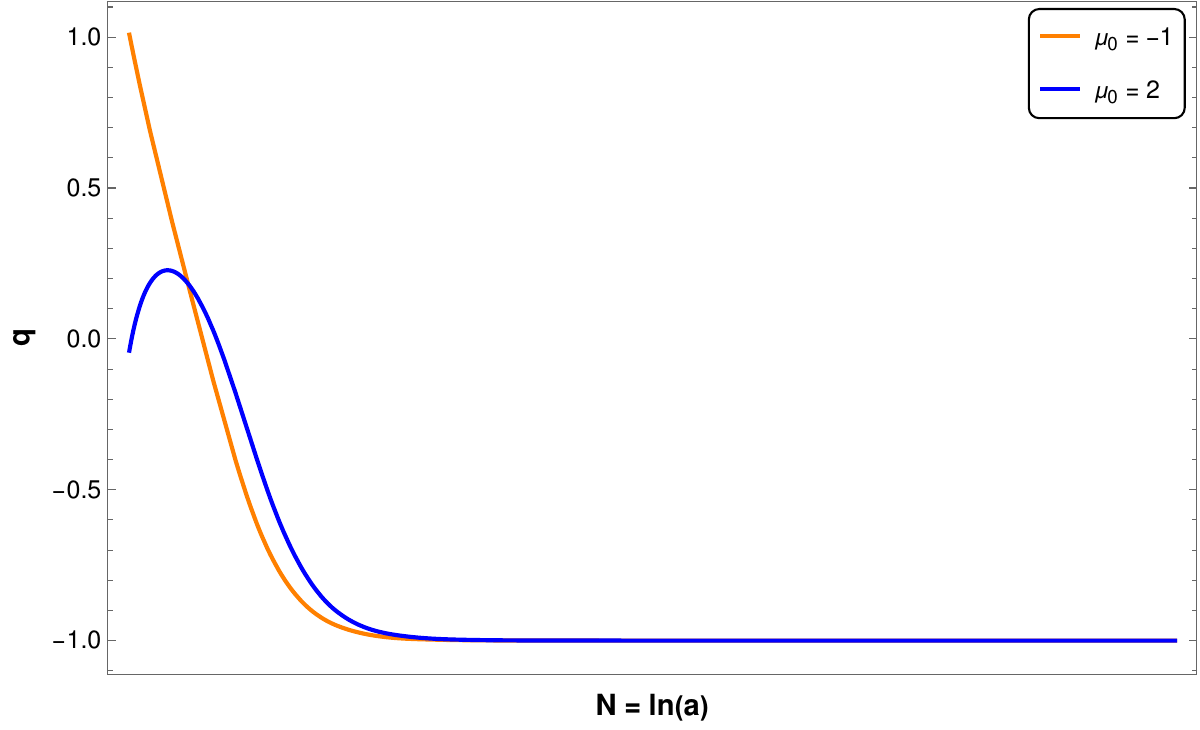}
    \end{subfigure}
    
    \caption{Qualitative evolution of the deceleration parameter for connection $\Gamma_B$ for different values of $\lambda_0$ and $\mu_0$, with initial conditions ($x_2[0]=0.1,~x_3[0]=0.1,~x_5[0]=0.1$) with parameter value is $h=0.5$.}
    \label{fig1_conll}
\end{figure}

For connection $\Gamma_B$, the qualitative evolution of the deceleration parameter for different parameters is shown in Fig \ref{fig1_conll}.
\begin{table}[h]
    \centering
      \begin{tabular}{|c|c|c|c|}
        \hline
        Critical point  &$(x_2,x_3,x_5)$ &Existence & $q$ \\ \hline
       $P_1$ & $\left( 0,0,2+\frac{2}{3(\mu_0-1)} \right)$& $\mu_0 \neq 1$&$\frac{1}{2}$ \\ \hline
         $P_2$ & $\left( -\frac{\sqrt{6}}{1+\lambda_0},0,2+\frac{6h-(\lambda_0-5)(\lambda_0+1)}{3(\lambda_0+1)(\mu_0-1)} \right)$ &  $\lambda_0 \neq -1 \land \mu_0 \neq 1$&$2-\frac{6}{1+\lambda_0}$\\ \hline
          $P_3$ & $\left(0,2,2-\frac{2(\lambda_0-2)}{3(\mu_0-1)} \right)$ & $\mu_0 \neq 1$&$-1$\\ \hline
          $P_4$ & $\left(x_2,0,\frac{\sqrt{\frac{2}{3}}(1-hx_2^2+\sqrt{6}x_2\mu_0)}{x_2(\mu_0-1)}\right)$ & $x_2 \neq 0\land\mu_0 \neq 1$&$2+\sqrt{6}x_2$\\ \hline
          \end{tabular}
    \caption{Critical points and their physical properties.}
    \label{table1_conll}
\end{table}

\begin{table}[h]
    \centering
    \begin{tabular}{|c|c|c|}
        \hline
        Critical point  & Eigenvalues & Stability \\ \hline
        $P_1$ & $\left(-\tfrac{3}{2},-\tfrac{3}{2}, 3 \right)$ & saddle \\ \hline
        $P_2$ & $\left( 0,0,3-\frac{6}{1+\lambda_0} \right)$ &unstable for $(\lambda_0<-1 \lor \lambda_0>1)$, otherwise non-hyperbolic \\ \hline
       $P_3$ & $\left( -3,-3,-3 \right)$ & stable \\ \hline
        $P_4$ & $\left( 0,6+\sqrt{6}x_2(1+\lambda_0),3+\sqrt{6}x_2 \right)$ &unstable for $(x_2 \in \mathbb{R}\land\lambda_0=-1 )$, otherwise non-hyperbolic \\ \hline
    \end{tabular}
    \caption{Eigenvalues and Stability.}
    \label{table2_conll}
\end{table}

\subsection{Phasespace and stability analysis of $\Gamma_C$}\label{sub3}
Under the assumptions (\ref{assumption}) and constraint equation (\ref{generic_constraint}), the three-dimensional dynamical system for this connection can be obtained by taking $x_5=0$ in equations (\ref{generic_eq1}) and (\ref{generic_eq4}). The detailed examination of the critical points is summarized in Table \ref{table1_conlll} and Table \ref{table2_conlll},  which specifies the conditions under which they exist, their related cosmological parameter $q$, and stability analysis. 

The critical point $P_1$ exists for $\mu_0 \neq 1$ and is characterized by a deceleration parameter $ q=\frac{1}{2}$, which corresponds to the standard dust-filled (matter-dominated) cosmological era. The eigenvalues $\{-\frac{3}{2},-\frac{1}{2},3\}$ indicate a saddle-type behavior, with one positive eigenvalue implying instability. Hence, $P_1$ represents a transient matter-dominated epoch, describing the intermediate stage between radiation and dark-energy domination.

The point $P_2\left(\frac{1}{\sqrt{6}},0,2+\frac{1}{1-\mu_0} \right)$ corresponds to a kinetic-dominated scalar field regime with a deceleration parameter $ q=1$, indicating a strongly decelerated universe. The instability condition $\lambda_0>-3$ shows that $P_2$ cannot serve as a late-time attractor but rather describes a stiff-matter or kinetic phase relevant to the early evolution of the universe.

The critical point $P_3$ exists for $\mu_0 \neq1$ and represents a pure de Sitter solution, where the expansion is exponentially accelerated with $ q=-1$. The eigenvalues of the linearized system are $\{-5,-3,-3\}$ are all negative, confirming that it is a stable attractor. Physically, this point corresponds to a late-time accelerated epoch, consistent with dark-energy–dominated cosmology.

The point $P_4$ exhibits the scaling solution. The deceleration parameter takes the form $q=\frac{2(\lambda_0-2)}{3\lambda_0-1}$. An accelerated expansion occurs when $(\frac{1}{3}<\lambda_0<2)$. The point remains stable within the parametric range ($-\frac{1}{7}<\lambda_0<\frac{1}{3} \land \left(\frac{-11+\lambda_0(2+3\lambda_0)}{2+14\lambda_0}<\mu_0<\frac{-59+3\lambda_0(8+\lambda_0)}{30+60\lambda_0}\right)$).

The critical point $P_5 \left(\frac{-2+3\mu_0}{\sqrt{6}(1+\mu_0)},0,0\right)$, existing for $\mu_0 \neq -1$, represents a solution where the dynamics are influenced by the parameter $\mu_0$. The deceleration parameter is $q=2-\frac{5}{2(1+\mu_0)}$ allowing cosmic acceleration when $(-1<\mu_0<\frac{1}{4})$. Stability is achieved for  $(\lambda_0<-3 \land \frac{-3+2\lambda_0}{3+3\lambda_0}<\mu_0<\frac{3}{2})$.

The point $P_6$ exists when $(2+\mu_0+\lambda_0\mu_0 \neq 0)$. The parametric dependent $q$ expression at this point is given as $q=\frac{\lambda_0^2-3+2\mu_0-4\lambda_0\mu_0}{2+\mu_0+\lambda_0\mu_0}$. This point is a stable for $(\frac{1}{3}<\lambda_0 \leq1 \land \mu_0<-\frac{2}{1+\lambda_0})$, while it supports accelerated expansion if $(-1<\mu_0<0 \land \lambda_0>\frac{-2-\mu_0}{\mu_0})$.

The critical points $P_{\pm7}$ also correspond to a solution in which the dynamics of the scalar field are also influenced by the parameter $\mu_0$. The point $P_{-7}$ exhibits a stable behavior for $(\mu_0<-\frac{1}{6} \land \lambda_0>1)$ and accounts for cosmic acceleration if $\mu_0<\frac{1}{6}$, on the other hand, the point $P_{+7}$ is always unstable and corresponds to a decelerating epoch, likely describing an early scalar dominated transient phase.

The point $P_{-8}$ exhibits stability for $(\mu_0<-1 \land \lambda_0 \leq-\frac{1}{2})$ and accounts for cosmic acceleration when $(-1<\mu_0<\frac{1}{6}$. But the point $P_{+8}$ always gives a decelerated universe with $(\mu_0<-1 \lor \mu_0>-1)$ with unstable nature when $\mu_0<-1$.
\begin{table}[h]
    \centering
      \begin{tabular}{|c|c|c|c|}
        \hline
        Critical point  &$(x_2,x_3,x_4)$ &Existence & $q$ \\ \hline
       $P_1$ & $\left( 0,0,2+\frac{2}{3(\mu_0-1)} \right)$& $\mu_0 \neq 1$&$\frac{1}{2}$ \\ \hline
         $P_2$ & $\left(\frac{1}{\sqrt{6}},0,2+\frac{1}{1-\mu_0} \right)$ &  $ \mu_0 \neq 1$&$1$\\ \hline
          $P_3$ & $\left(0,2,2-\frac{2(\lambda_0-2)}{3(\mu_0-1)} \right)$ & $\mu_0 \neq 1$&$-1$\\ \hline
           $P_4$ & $\left(\frac{5\sqrt{\frac{2}{3}}}{1-3\lambda_0},\frac{4(59-3\lambda_0(8+3\lambda_0-20\mu_0)+30\mu_0)}{3(1-3\lambda_0)^2},\frac{-11+\lambda_0(2+3\lambda_0-14\mu_0)-2\mu_0}{(-1+\mu_0)(-1+3\lambda_0)}\right)$ & $\lambda_0 \neq \frac{1}{3}\land\mu_0 \neq 1$&$\frac{2(\lambda_0-2)}{3\lambda_0-1}$\\ \hline
            $P_5$ & $\left(\frac{-2+3\mu_0}{\sqrt{6}(1+\mu_0)},0,0\right)$ & $\mu_0 \neq -1$&$2-\frac{5}{2(1+\mu_0)}$\\ \hline
             $P_6$ & $\left(\frac{\sqrt{\frac{2}{3}}(3\mu_0-1-\lambda_0)}{2+\mu_0+\lambda_0 \mu_0},-\frac{2(-5+2\lambda_0+\lambda_0^2-6(1+\lambda_0)\mu_0)(2+3\mu_0^2)}{3(2+\mu_0+\lambda_0\mu_0)^2},0\right)$ & $2+\mu_0+\lambda_0\mu_0 \neq 0$&$\frac{\lambda_0^2-3+2\mu_0-4\lambda_0\mu_0}{2+\mu_0+\lambda_0\mu_0}$\\ \hline
              $P_{\pm7}$ & $\left(\frac{\sqrt{3}\mu_0 \pm\sqrt{2+3\mu_0^2}}{\sqrt{2}},0,0\right)$ &Always&$2+3\mu_0\pm\sqrt{6+9\mu_0^2}$\\ \hline
              $P_{\pm8}$ & $\left(\frac{-1+2\mu_0 \pm \sqrt{3-2\mu_0+4\mu_0^2}}{\sqrt{6}(1+\mu_0)},0,\frac{4+\mu_0\pm2\sqrt{3-2\mu_0+4\mu_0^2}\pm3\mu_0\sqrt{3-2\mu_0+4\mu_0^2}}{3-3\mu_0^2}\right)$ & $\mu_0 \neq \pm1$&$\frac{1+4\mu_0\pm \sqrt{3-2\mu_0+4\mu_0^2}}{3+3\mu_0}$\\ \hline
          \end{tabular}
    \caption{Critical points and their physical properties.}
    \label{table1_conlll}
\end{table}

\begin{table}[H]
\centering
\scriptsize
\begin{tabular}{|c|c|c|}
\hline
 Critical point  & Eigenvalues & Stability \\ \hline
 $P_{1}$ & $\left(-\tfrac{3}{2},-\tfrac{1}{2}, 3 \right)$ & saddle \\ \hline
 $P_{2}$ & $\left( 3+\lambda_0,A_+,A_-\right)$   & unstable for $ \lambda_0>-3 $\\ \hline
 $P_{3}$ & $\left( -5,-3,-3 \right)$ & stable \\ \hline
$P_{4}$ & $\left(\frac{3+\lambda_0}{3\lambda_0-1}, C_+,C_-\right)$ &\begin{tabular}[c]{@{}l@{}}%
stable for $-\frac{1}{7}<\lambda_0<\frac{1}{3} \land $\\
$\left(\frac{-11+\lambda_0(2+3\lambda_0)}{2+14\lambda_0}<\mu_0<\frac{-59+3\lambda_0(8+\lambda_0)}{30+60\lambda_0}\right)$%
\end{tabular} \\ \hline
 $P_{5}$ & $\left( \frac{1}{2}-\frac{5}{4(1+\mu_0)},-3+\frac{5}{2(1+\mu_0)}, 3+\lambda_0(3-\frac{5}{(1+\mu_0)})\right)$ & stable for $\lambda_0<-3 \land \frac{-3+2\lambda_0}{3+3\lambda_0}<\mu_0<\frac{3}{2}$ \\ \hline
 $P_{6}$ & $\left( \frac{-11+\lambda_0(2+3\lambda_0-14\mu_0)-2\mu_0}{2(2+\mu_0+\lambda_0\mu_0)},\frac{\lambda_0(2+2\lambda_0-9\mu_0)-3(2+\mu_0)}{(2+\mu_0+\lambda_0\mu_0)}, -6+\frac{7+\lambda_0(2+\lambda_0)}{(2+\mu_0+\lambda_0\mu_0)}\right)$ &stable for $\frac{1}{3}<\lambda_0 \leq1 \land \mu_0<-\frac{2}{1+\lambda_0}$ \\ \hline
        $P_{\pm7}$ & $\left((3+3\mu_0\pm\sqrt{6+9\mu_0^2},6+3(1+\lambda_0)\mu_0\pm\sqrt{6+9\mu_0^2}\pm \lambda_0\sqrt{6+9\mu_0^2},2+3\mu_0\pm\sqrt{6+9\mu_0^2})\right)$& $unstable_{+}$, stable for $[\mu_0<-\frac{1}{6} \land \lambda_0>1]_{-}$\\\hline
        $P_{\pm 8}$ &
$\left(E_{1\pm},E_{2\pm},E_{3\pm}\right)$ &
\begin{tabular}[c]{@{}l@{}}%
unstable for $[\mu_0<-1]_{+}$,\\
stable for $[\mu_0<-1 \land \lambda_0 \leq-\frac{1}{2}]_{-}$%
\end{tabular} \\ \hline
    \end{tabular}
    \caption{Eigenvalues and Stability} \caption*{\footnotesize Due to the complexity of some eigenvalue expressions, they are not explicitly presented in the table and are instead denoted by capital letters for brevity. }.
    \label{table2_conlll}
\end{table}

\begin{figure}[tbp]
    \centering
    \begin{subfigure}[b]{0.45\textwidth}
        \centering
        \includegraphics[width=\textwidth]{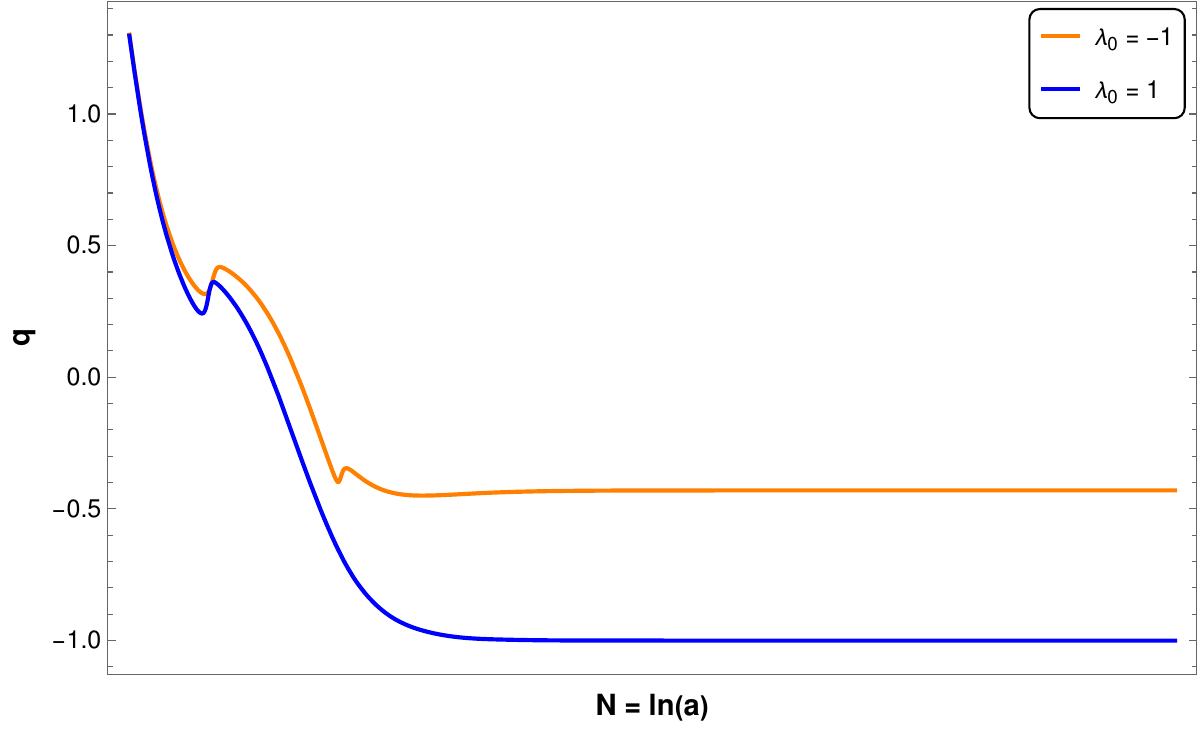}
    \end{subfigure}
    \begin{subfigure}[b]{0.45\textwidth}
        \centering
        \includegraphics[width=\textwidth]{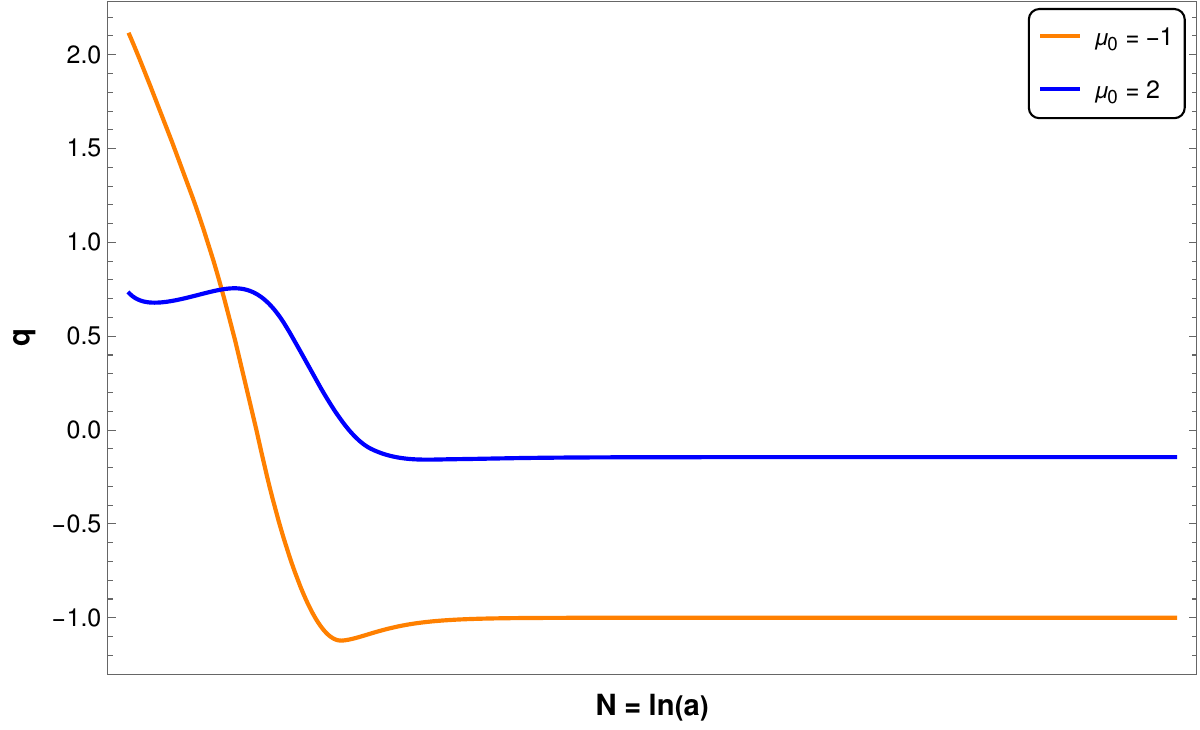}
    \end{subfigure}
    
    \caption{Qualitative evolution of the deceleration parameter for connection $\Gamma_C$
for different values of $\lambda_0$ and $\mu_0$, with initial conditions ($x_2[0]=0.8,~x_3[0]=0.01,~x_4[0]=0.05$) with parameter value is $h=0.5$.}
    \label{fig1_conlll}
\end{figure}

\section{Conclusions}\label{conlusion}
We have proposed and analyzed a novel symmetric teleparallel gravity in which a single scalar field is non-minimally coupled to the non-metricity scalar $Q$ and the associated boundary term $C$. This construction extends scalar-tensor gravity beyond metric compatibility and goes beyond minimal $f(Q)$ scenarios by activating boundary interactions that reshuffle the effective dynamics. The covariant formulation, energy conservation, and cosmological implications of the theory have been presented. We have also discussed conformal transformation and remarkably, the theory admits an Einstein frame formulation dynamically equivalent to STEGR, a property that is absent in scalar-tensor theories based solely on non-metricity. Another central outcome of this work is a unified autonomous system setup that accommodates generic affine-connection classes (including the coincident gauge and non-coincident branches) within a single dynamical framework, providing a powerful and systematic framework for the analysis of cosmological dynamics in non-metricity based gravity.

From the dynamical systems perspective, we choose $\zeta,\mu $ and $\lambda$, which are parameters corresponding to the coupling functions $f$, $\chi$ and potential $U$, to be constants and obtain a power law form of these functions which we use to rewrite the autonomous dynamical system for unified formulation and also across all three connections. We identify CPs, their existence conditions, determine their stability through linearization around each equilibrium, and calculate the corresponding deceleration parameter value for each critical point.

For the connection $\Gamma_A$, the two dimensional dynamical system  yields four critical points. Among these, the critical points $P_1$ and $P_{\pm3}$ represent solutions where the evolution of the scalar field is governed by the parameters $\mu_0$ and $h$ and can account for the cosmic acceleration under some particular parametric choice. The critical point $P_2$ describes a scaling solution and can also describe an accelerating universe under specific parametric constraints.

For the connection $\Gamma_B$, the three dimensional dynamical system admits four critical points. The critical point $P_1$ corresponds to a transient, matter-dominated phase of the universe, whereas $P_3$ represents a stable late-time de Sitter attractor solution. Meanwhile, $P_2$ and $P_4$ can describe an accelerated expansion of the universe within specific parameter ranges. 

For the connection $\Gamma_C$, the corresponding three-dimensional dynamical system possesses ten distinct critical points. The critical point $P_1$ represents a dust-dominated cosmological phase, while $P_2$ describes a decelerating epoch associated with the early universe. The point $P_3$ represents a stable late-time de Sitter attractor. Furthermore, the critical points $P_4$, $P_5$, $P_6$, $P_{-7}$, and $P_{-8}$ can lead to an accelerated expansion of the universe under specific parameter constraints, whereas $P_{+7}$ and $P_{+8}$ describe a decelerated cosmological phase.

The present results establish scalar non-metricity gravity with boundary term couplings as a flexible, geometrically motivated framework in which cosmic acceleration and a standard expansion history emerge as robust dynamical outcomes. Future work should extend the analysis beyond the background: (i) a full perturbation study for the generic affine branch to assess stability, effective Newton’s constant, and slip at sub-horizon scales; (ii) reconstruction strategies for $f(\phi)$, $\chi(\phi)$, and $U(\phi)$ from background and growth data; and (iii) confrontation with multi-messenger bounds, ensuring scalar non-metricity braiding remains consistent with gravitational-wave propagation.

\section{Appendix}
\appendix
\section{Equations of motion of $\Gamma_A$}\label{GammaA}
In the class of affine connections $\Gamma_A$, we have  $C_1=\gamma$; $C_2=C_3=0$ and $k=0$, and we can deduce the following equations of motion.
\begin{align}\label{Q_con1}
Q=&-6H^2\,.
\end{align}
\begin{align}\label{p_con1}
\kappa p=(-2\dot H-3H^2)f-2H(\dot f-\dot\chi)+\frac U2-\frac{h\dot \phi^2}{2}-2H\dot\chi-\ddot\chi
\end{align}
\begin{align}\label{rho_con1}
\kappa\rho=3H^2f-\frac U2-\frac{h\dot \phi^2}{2}+3H\dot\chi.
\end{align}
The scalar field Eq can be given as
\begin{align}\label{sclreq_con1}
f'(-6H^2)+\chi'\left(6\dot{H}+18H^2\right)-2h(\ddot\phi+3H\dot\phi)-U'=0.
\end{align}
\begin{align}
\kappa\dot{\rho}+3\kappa H(\rho+p)=0
\end{align}
\section{Equations of motion of $\Gamma_B$}\label{GammaB}
In this class of connections, we have the following.
$C_1=\gamma+\dfrac{\dot \gamma}\gamma$; $C_2=0$; $C_3=\gamma$ and $k=0$.
\begin{align}
Q=&3\left(-2H^2+3H\gamma+\dot \gamma\right)\,.
\end{align}
\begin{align}\label{p_conii}
    \kappa p=(-2\dot H-3H^2)f+\frac{3\gamma-4H}2(\dot f-\dot\chi)+\frac U2-\frac{h\dot \phi^2}{2}-2H\dot\chi-\ddot\chi
\end{align}
\begin{align}\label{rho_eqconii}
    \kappa\rho=3H^2f+\frac{3\gamma}2(\dot f-\dot\chi)-\frac U2-\frac{h\dot \phi^2}{2}+3H\dot\chi.
\end{align}

The scalar field Eq can be written as
\begin{align}\label{scalareq_conii}
f'\left(-6H^2+9H\gamma+3\dot \gamma\right)+\chi'\left(6\dot{H}+18H^2-9H\gamma-3\dot \gamma\right)-2h(\ddot\phi+3H\dot\phi)-U'=0.
\end{align}
\begin{align}\label{conectioneq_conii}
\kappa\dot{\rho}+3\kappa H(\rho+p)
=\frac32\gamma\left[  3(\dot f-\dot\chi) H +(\ddot f-\ddot\chi) \right].
\end{align}

\section{Equations of motion of $\Gamma_C$}\label{GammaC}
In this class, we have the following.
$C_1=-\dfrac k\gamma-\dfrac{\dot \gamma}\gamma$; $C_2=\gamma$; 
$C_3=-\dfrac k\gamma$ and $k=0,\pm1$.
\begin{align}
Q=&3\left(-2H^2+2\frac k{a^2}-\frac{3kH}\gamma+\frac{k\dot\gamma}{\gamma^2}+\frac1{a^2}(H\gamma+\dot \gamma)\right)\,.
\end{align}
\begin{align}\label{p_coniii}
\kappa p
=f\left(-2\dot H-3H^2-\frac k{a^2}\right)
    +\frac12(\dot f-\dot \chi) \left(-\frac{3k}{\gamma}+\frac{\gamma}{a^2}-4H\right)
    -\frac12h\dot \phi^2    +\frac12U -2H\dot \chi-\ddot\chi \,,
\end{align}
\begin{align}\label{rho_coniii}
\kappa\rho
=f\left(3H^2+3\dfrac k{a^2}\right)
    +\frac12(\dot f-\dot\chi) \left(-\frac{3k}{\gamma}-3\frac{\gamma}{a^2}\right)
     -\frac12h\dot \phi^2 -\frac12U+3H\dot \chi \,,
\end{align}

The scalar field Eq can be expressed as
\begin{align}\label{scalareq_coniii}
&f'\left(-6H^2+6\frac k{a^2}-\frac{9kH}\gamma+\frac{3k\dot\gamma}{\gamma^2}+\frac3{a^2}(H\gamma+\dot \gamma)\right)+\chi'\left(6\dot{H}+18H^2-\frac{6k}{a^2}+\frac{9Hk}{\gamma}-\frac{3k\dot{ \gamma}}{\gamma^2}-\frac{3}{a^2}(H\gamma+\dot{\gamma})\right)\notag\\&-2h(\ddot\phi+3H\dot\phi)-U'=0.
\end{align}
\begin{align}\label{conectioneq_coniii}
\kappa\dot{\rho}+3\kappa H(\rho+p)
=-\frac32\left[
    (\dot f-\dot\chi)\left(3\frac k\gamma H+\frac{2\dot \gamma+\gamma H}{a^2}\right)
    +(\ddot f-\ddot\chi)\left(\frac k\gamma+\frac{\gamma}{a^2}\right)\right].
\end{align}
\section{The autonomous system components}\label{autom}
\begin{align}\label{gen_eqfirst}
E_1= -\Bigg(&
3 x_1 \Big(
-3(\mu-\zeta)\, x_3 (x_4-x_5)^2
+ \sqrt{6}\, x_2 \Big(
(\mu-\zeta)^2 x_4^3
+ (\mu-\zeta) x_4^2 \big(2(\mu+\zeta)+(\mu-\zeta)x_5\big)
\notag\\
&
+ (\mu-\zeta)x_5^2 \big(-6\mu+2\zeta+3(\mu-\zeta)x_5\big)
+ x_4 \Big(
4\mu(-3\mu+2\zeta)
+4(h+\lambda\mu)x_3
\notag\\
&
-(\mu-\zeta)x_5\big(4(-7\mu+\zeta)+5(\mu-\zeta)x_5\big)
\Big)
\Big)
\Big)
-8\sqrt{6}\, h\, x_2^3 x_4 \Big(h+2\eta^2\Theta\Big)
\notag\\
&
-2 x_2^2 \Big(
3x_5^2\Big(h(-\mu+\zeta)-2\mu\zeta^2\Delta+2\zeta \eta^2\Theta\Big)
+x_4^2\Big(h(\mu-\zeta)-6\mu \zeta^2\Delta+6\zeta \eta^2\Theta\Big)
+2x_4\Big(
-6\mu^2\zeta
+4h(-3\mu+\zeta)
\notag\\
&
+x_5\big(
9h\mu-9h\zeta
+6\mu \zeta^2\Delta
-6\zeta \eta^2\Theta
\big)
\Big)
\Big)
\Bigg) \Bigg/ \Big(4 \sqrt{6}
  \,x_2(2h +3 \mu^2)x_4 
  -6(\mu-\zeta )(x_4 - x_5)^2
\Big)\Bigg)
\end{align}

\begin{align}
E_2=
-\Bigg(& 3 x_2 \Big(
\sqrt{6} (\mu - \zeta)^2 x_2 x_4^3 
+ x_4^2 \Big(-(\mu - \zeta)(-10 + 3x_3) 
+ \sqrt{6} x_2 \big(2(\mu - \zeta)(3\mu + \zeta) 
+ (\mu - \zeta)^2 x_5 
+ 4 \zeta^2\Delta - 4 \eta^2\Theta\big) \notag\\
& + 2x_2^2(-h\mu + h\zeta + 6\mu \zeta^2\Delta - 6\zeta \eta^2\Theta)\Big) + x_5^2 \Big(-3(\mu - \zeta)(2 + x_3) 
+ 6x_2^2(h\mu - h\zeta + 2\mu \zeta^2\Delta - 2\zeta \eta^2\Theta) \notag\\
&+ \sqrt{6}x_2 \big(3(\mu - \zeta)^2 x_5- 2(3\mu^2 - 4\mu\zeta + \zeta^2 - 2\zeta^2\Delta + 2\eta^2\Theta)\big)\Big) \notag\\
&+ x_4 \Big(4(-6\mu + 4\zeta + (2\lambda - 3\mu)x_3) 
+ 2(\mu - \zeta)(22 + 3x_3)x_5 
- 8\sqrt{6}h x_2^3 (h + 2\eta^2\Theta)  \notag\\
& - 4x_2^2\Big(-2(9h\mu - 2h\zeta + 3\mu^2\zeta + 6\mu \eta^2\Theta) 
+ x_5(9h\mu - 9h\zeta + 6\mu \zeta^2\Delta - 6\zeta \eta^2\Theta)\Big)  \notag\\
& + \sqrt{6}x_2\Big(8(h + 3\mu(-\mu + \zeta)) 
+ 4(h + \lambda\mu)x_3  \notag\\
&+ x_5\big(-5(\mu - \zeta)^2 x_5 
+ 4(10\mu^2 - 11\mu\zeta + \zeta^2 - 2\zeta^2\Delta + 2\eta^2\Theta)\big)\Big)
\Big)
\Big)\Bigg/ 4 \Big( 
  2\sqrt{6}\,x_2(2h + 3\mu^2)x_4 
  + 3(\zeta - \mu)(x_4 - x_5)^2
\Big)\Bigg)
\end{align}

\begin{align}
E_3=
-\Bigg(&
3 x_3 \Big(
-3 (\mu - \zeta)(-2 + x_3)(x_4 - x_5)^2 
+ \sqrt{6} x_2 \Big((\mu - \zeta)^2 x_4^3 
+ (\mu - \zeta) x_4^2 \big(2(\lambda + \mu + \zeta) + (\mu - \zeta)x_5\big)\notag \\
&  + (\mu - \zeta) x_5^2 \big(2(\lambda - 3\mu + \zeta) + 3(\mu - \zeta)x_5\big) 
+ x_4 \big(-8(h + \mu(3\mu - \zeta)) + 4(h + \lambda\mu)x_3\notag \\
&  + (\mu - \zeta)x_5 \big(-4(\lambda - 7\mu + \zeta) + 5(-\mu + \zeta)x_5\big)\big)\Big) \notag \\
&  - 8\sqrt{6}h x_2^3 x_4 (h + 2\eta^2\Theta) 
- 2x_2^2 \Big(3x_5^2\big(h(-\mu + \zeta) - 2\mu \zeta^2\Delta + 2\zeta \eta^2\Theta\big) \notag \notag \\
&   + x_4^2 \big(h(\mu - \zeta) - 6\mu \zeta^2\Delta + 6\zeta \eta^2\Theta\big) 
+ 2x_4 \big(6\mu^2(\lambda - \zeta) + 4h(\lambda - 3\mu + \zeta)\notag \\
&  + x_5(9h\mu - 9h\zeta + 6\mu \zeta^2\Delta - 6\zeta \eta^2\Theta)\big)\Big)
\Big)\Bigg/ 4\sqrt{6}\,x_2(2h + 3\mu^2)x_4 
+ 6(\zeta - \mu)(x_4 - x_5)^2\Bigg)
\end{align}

\begin{align}
E_4=-\Bigg(& x_4 \Big(
3\sqrt{6}(\mu - \zeta)^3 x_2 x_4^3 
+ 3(\mu - \zeta)x_4^2 \Big(-3(\mu - \zeta)(2 + x_3)
+ \sqrt{6}(\mu - \zeta)x_2\big(4\zeta + (\mu - \zeta)x_5\big) \notag \\
&  + 2x_2^2\big(-h\mu + h\zeta + 6\mu \zeta^2\Delta - 6\zeta \eta^2\Theta\big)\Big)  + x_4 \Big(6(\mu - \zeta)\big(6\mu - 4\zeta + (-2\lambda + 3\mu)x_3 
+ 3(\mu - \zeta)(-2 + x_3)x_5\big) \notag \\
&  + \sqrt{6}(\mu - \zeta)x_2\Big(-4(2h + 3\mu^2)
+ 12(h + \lambda\mu)x_3 
- 3(\mu - \zeta)x_5\big(8(-3\mu + \zeta) + 5(\mu - \zeta)x_5\big)\Big)\notag \\
&  - 24\sqrt{6}h(\mu - \zeta)x_2^3(h + 2\eta^2\Theta) 
+ 12x_2^2\Big(h(9\mu - 8\zeta)(\mu - \zeta) 
- 2(2h + 3\mu^2)\zeta^2\Delta + 2(2h + 3\mu\zeta)\eta^2\Theta\notag \\
&  - 3(\mu - \zeta)x_5(3h\mu - 3h\zeta + 2\mu \zeta^2\Delta - 2\zeta \eta^2\Theta)\Big)\Big)\notag + 3x_5 \Big(\sqrt{6}(\mu - \zeta)x_2(2\mu + (\mu - \zeta)x_5)(-6\mu + 4\zeta + 3(\mu - \zeta)x_5)\notag \\
&  + (\mu - \zeta)\big(-12\mu + 8\zeta + (4\lambda - 6\mu)x_3 
- 3(\mu - \zeta)(-6 + x_3)x_5\big)\notag \\
& 
+ 2x_2^2\big(6h\mu(\mu - \zeta) + 4(2h + 3\mu^2)\zeta^2\Delta - 4(2h + 3\mu\zeta)\eta^2\Theta\notag \\
&  
+ 3(\mu - \zeta)x_5(h\mu - h\zeta + 2\mu \zeta^2\Delta - 2\zeta \eta^2\Theta)\big)\Big)
\Big)\Bigg/ 
\Big(4(-\zeta + \mu)
\left(
  2\sqrt{6}\,x_2(2h + 3\mu^2)x_4
  + 3(\zeta - \mu)(x_4 - x_5)^2
\right)\Big)\Bigg)
\end{align}

\begin{align}
E_5=-\Bigg(& x_{5} \Big( 
  3\sqrt{6} (\mu - \zeta)^3 x_{2} x_{4}^3 
 + 3 (\mu - \zeta) x_{4}^2 
 \Big(
    -3(\mu - \zeta)(-6 + x_{3}) 
    + \sqrt{6} (\mu - \zeta) x_{2} 
      \big( 4(\mu + \zeta) + (\mu - \zeta)x_{5} \big)  \notag \\
&  + 2 x_{2}^2 \big( -h\mu + h\zeta + 6\mu \zeta^2\Delta - 6\zeta \eta^2\Theta \big)
 \Big)  - x_{4} \Big(
     6 (\mu - \zeta)
        \big( 6\mu - 4\zeta + (-2\lambda + 3\mu)x_{3} 
             - 3(\mu - \zeta)(-2 + x_{3})x_{5} \big)  \notag \\
&  + \sqrt{6} (\mu - \zeta) x_{2}
       \Big( 40h + 132\mu^2 - 48\mu\zeta 
           - 12(h + \lambda\mu)x_{3}
           + 3(\mu - \zeta)x_{5}\big(8(-4\mu + \zeta) + 5(\mu - \zeta)x_{5}\big)
       \Big)  \notag \\
&  + 24\sqrt{6}\, h(\mu - \zeta)x_{2}^3 \big(h + 2\eta^2\Theta\big)    + 12x_{2}^2 
       \Big(
           -h(15\mu - 8\zeta)(\mu - \zeta)
           - 2(2h + 3\mu^2) \zeta^2\Delta
           + 2(2h + 3\mu\zeta)\eta^2\Theta \notag \\
&
           + 3(\mu - \zeta)x_{5}
              (3h\mu - 3h\zeta + 2\mu \zeta^2\Delta - 2\zeta \eta^2\Theta)
       \Big)
 \Big)  + 3x_{5} 
 \Big(
     \sqrt{6}(\mu - \zeta)x_{2}(-2\mu + (\mu - \zeta)x_{5})(-6\mu + 4\zeta + 3(\mu - \zeta)x_{5}) \notag \\
&  + (\mu - \zeta)\big(12\mu - 8\zeta + (-4\lambda + 6\mu)x_{3} - 3(\mu - \zeta)(2 + x_{3})x_{5}\big)\notag \\
&  + 2x_{2}^2 
       \Big(
           6h\mu(-\mu + \zeta)
           - 4(2h + 3\mu^2)\zeta^2\Delta
           + 4(2h + 3\mu\zeta) \eta^2\Theta\notag \\
& 
           + 3(\mu - \zeta)x_{5}(h\mu - h\zeta + 2\mu \zeta^2\Delta - 2\zeta \eta^2\Theta)
       \Big)
 \Big)
\Big)\Bigg/ 
\Big(4(-\zeta + \mu)
\left(
  2\sqrt{6}\,x_2(2h + 3\mu^2)x_4
  + 3(\zeta - \mu)(x_4 - x_5)^2
\right)\Big)\Bigg)
\end{align}
\begin{align}
E_6=\sqrt{\tfrac{3}{2}}\,x_2\,
\big(-\zeta \mu + 2\eta^2\Theta\big)
\end{align}

\begin{align}
E_7=x_2\left(\frac{3\lambda(-2\lambda + \zeta)}{\sqrt{6}}+\sqrt{6}\Gamma\lambda^2\right)
\end{align}
\begin{align}
E_8=- \sqrt{\frac{3}{2}}\, x_{2}
\left(
\zeta^{2} - 2 \zeta^2\Delta
\right)
\end{align}
\begin{align}\label{gen_eqlast}
E_9=\frac{\sqrt{6}x_2\eta^3}{\mu}
\left(\Theta-1
\right)
\end{align}

The Eq for $\frac{\dot{H}}{H^2}$ can be written as
\begin{align}
\frac{\dot{H}}{H^2}=3 \Bigg(&
-3 (\mu - \zeta) (-2 + x_3) (x_4 - x_5)^2  + \sqrt{6}\, x_2 \Big(
(\mu -  \zeta)^2 x_4^3 
+ (\mu -  \zeta) x_4^2 \left( 2 (\mu + 2  \zeta) + (\mu -  \zeta) x_5 \right)  \nonumber\\
&+ (\mu -  \zeta) x_5^2 \left( -6 \mu + 4  \zeta + 3 (\mu -  \zeta) x_5 \right)+ x_4 \Big(
-8 (h + \mu (3 \mu -  \zeta))
+ 4 (h + \lambda \mu) x_3  \nonumber\\
& + (\mu -  \zeta) x_5
\left( 28 \mu - 8  \zeta + 5 (-\mu +  \zeta) x_5 \right)
\Big)
\Big) - 8 \sqrt{6}\, h\, x_2^3 x_4 
\left( h + 2 \eta^2 \Theta \right) \nonumber\\
&- 2 x_2^2 \Big(
3 x_5^2 \left(
- h \mu + h  \zeta
- 2 \mu \zeta^2\Delta
+ 2  \zeta \eta^2 \Theta
\right) + x_4^2 \left(
h \mu - h  \zeta
- 6 \mu \zeta^2\Delta
+ 6  \zeta \eta^2 \Theta
\right)  \nonumber\\
& + 2 x_4 \Big(
-12 h \mu + 8 h  \zeta
+ x_5 \big(
9 h \mu - 9 h  \zeta
+ 6 \mu \zeta^2\Delta
- 6  \zeta \eta^2 \Theta
\big)
\Big)
\Big)
\Bigg)\Bigg/ 
\left(
  8\sqrt{6}\,x_2(2h + 3\mu^2)x_4
  -12(\mu-\zeta)(x_4 - x_5)^2
\right).
\end{align}

Under the particular choices of coupling and potential functions, and using the constraint equation, these reduce to
\begin{align}\label{generic_eq1}
\mathbb{E}_2=-\Bigg(& 3 x_2 \Big(
\sqrt{6} (\mu_0 - 1)^2 x_2 x_4^3 
+ x_4^2 \Big(-(\mu_0 - 1)(-10 + 3x_3) 
+ \sqrt{6} x_2 \big(2(\mu_0 - 1)(3\mu_0 + 1) 
+ (\mu_0 - 1)^2 x_5 
\big) \notag\\
& + 2x_2^2(-h\mu_0 + h + 3\mu_0  - 3)\Big) + x_5^2 \Big(-3(\mu_0 - 1)(2 + x_3) 
+ 6x_2^2(h\mu_0 - h + 2\mu_0  - 1) \notag\\
&+ \sqrt{6}x_2 \big(3(\mu_0 - 1)^2 x_5- 2(3\mu_0^2 - 4\mu_0  + 1)\big)\Big) \notag\\
&+ x_4 \Big(4(-6\mu_0 + 4+ (2\lambda_0 - 3\mu_0)x_3) 
+ 2(\mu_0 - 1)(22 + 3x_3)x_5 
- 8\sqrt{6}h x_2^3 (h +1)  \notag\\
& - 4x_2^2\Big(-2(9h\mu_0 - 2h + 3\mu_0^2+ 3\mu_0 ) 
+ x_5(9h\mu_0 - 9h + 3\mu_0  - 3)\Big)  \notag\\
& + \sqrt{6}x_2\Big(8(h + 3\mu_0(-\mu_0 + 1)) 
+ 4(h + \lambda_0\mu_0)x_3  \notag\\
&+ x_5\big(-5(\mu_0 -1)^2 x_5 
+ 4(10\mu_0^2 - 11\mu_0  +1)\big)\Big)
\Big)
\Big)\Bigg/ 4 \Big( 
  2\sqrt{6}\,x_2(2h + 3\mu_0^2)x_4 
  + 3(1 - \mu_0)(x_4 - x_5)^2
\Big)\Bigg)
\end{align}
\begin{align}
\mathbb{E}_3=-\Bigg(&
3 x_3 \Big(
-3 (\mu_0 - 1)(-2 + x_3)(x_4 - x_5)^2 
+ \sqrt{6} x_2 \Big((\mu_0 -1)^2 x_4^3 
+ (\mu_0 - 1) x_4^2 \big(2(\lambda_0 + \mu_0 + 1) + (\mu_0 - 1)x_5\big)\notag \\
&  + (\mu_0 - 1) x_5^2 \big(2(\lambda_0 - 3\mu_0 + 1) + 3(\mu_0 - 1)x_5\big) 
+ x_4 \big(-8(h + \mu_0(3\mu_0 - 1)) + 4(h + \lambda_0\mu_0)x_3\notag \\
&  + (\mu_0 - 1)x_5 \big(-4(\lambda_0 - 7\mu_0+ 1) + 5(-\mu_0 + 1)x_5\big)\big)\Big) \notag \\
&  - 8\sqrt{6}h x_2^3 x_4 (h +1) 
- 2x_2^2 \Big(3x_5^2\big(h(-\mu_0 + 1) - \mu_0  +1\big) \notag \notag \\
&   + x_4^2 \big(h(\mu_0 - 1) - 3\mu_0  +3\big) 
+ 2x_4 \big(6\mu_0^2(\lambda_0 - 1) + 4h(\lambda_0 - 3\mu_0 + 1)\notag \\
&  + x_5(9h\mu_0 - 9h + 3\mu_0  - 3)\big)\Big)
\Big)\Bigg/ 4\sqrt{6}\,x_2(2h + 3\mu_0^2)x_4 
+ 6(1 - \mu_0)(x_4 - x_5)^2\Bigg)
\end{align}

\begin{align}
\mathbb{E}_4=-\Bigg(& x_4 \Big(
3\sqrt{6}(\mu_0 -1)^3 x_2 x_4^3 
+ 3(\mu_0 - 1)x_4^2 \Big(-3(\mu_0 - 1)(2 + x_3)
+ \sqrt{6}(\mu_0 - 1)x_2\big(4 + (\mu_0 - 1)x_5\big) \notag \\
&  + 2x_2^2\big(-h\mu_0 + h + 3\mu_0  -3\big)\Big)  + x_4 \Big(6(\mu_0 - 1)\big(6\mu_0 - 4+ (-2\lambda_0 + 3\mu_0)x_3 
+ 3(\mu_0 - 1)(-2 + x_3)x_5\big) \notag \\
&  + \sqrt{6}(\mu_0 - 1)x_2\Big(-4(2h + 3\mu_0^2)
+ 12(h + \lambda_0\mu_0)x_3 
- 3(\mu_0 - 1)x_5\big(8(-3\mu_0 + 1) + 5(\mu_0 - 1)x_5\big)\Big)\notag \\
&  - 24\sqrt{6}h(\mu_0 - 1)x_2^3(h +1) 
+ 12x_2^2\Big(h(9\mu_0 - 8)(\mu_0 - 1) 
- (2h + 3\mu_0^2) + (2h + 3\mu_0)\notag \\
&  - 3(\mu_0 - 1)x_5(3h\mu_0 - 3h+ \mu_0  - 1)\Big)\Big)\notag + 3x_5 \Big(\sqrt{6}(\mu_0 - 1)x_2(2\mu_0 + (\mu_0 - 1)x_5)(-6\mu_0 + 4 + 3(\mu_0 -1)x_5)\notag \\
&  + (\mu_0 - 1)\big(-12\mu_0 + 8 + (4\lambda_0 - 6\mu_0)x_3 
- 3(\mu_0 -1)(-6 + x_3)x_5\big)\notag \\
& 
+ 2x_2^2\big(6h\mu_0(\mu_0 -1) + 2(2h + 3\mu_0^2) - 2(2h + 3\mu_0)\notag \\
&  
+ 3(\mu_0 - 1)x_5(h\mu_0 - h + \mu_0  -1)\big)\Big)
\Big)\Bigg/ 
\Big(4(-1+ \mu_0)
\left(
  2\sqrt{6}\,x_2(2h + 3\mu_0^2)x_4
  + 3(1 - \mu_0)(x_4 - x_5)^2
\right)\Big)\Bigg)
\end{align}

\begin{align}\label{generic_eq4}
\mathbb{E}_5=-\Bigg(& x_{5} \Big( 
  3\sqrt{6} (\mu_0 - 1)^3 x_{2} x_{4}^3 
 + 3 (\mu_0 - 1) x_{4}^2 
 \Big(
    -3(\mu_0 -1)(-6 + x_{3}) 
    + \sqrt{6} (\mu_0 - 1) x_{2} 
      \big( 4(\mu_0 + 1) + (\mu_0 - 1)x_{5} \big)  \notag \\
&  + 2 x_{2}^2 \big( -h\mu_0 + h + 3\mu_0  -3 \big)
 \Big)  - x_{4} \Big(
     6 (\mu_0 - 1)
        \big( 6\mu_0 - 4 + (-2\lambda_0 + 3\mu_0)x_{3} 
             - 3(\mu_0 - 1)(-2 + x_{3})x_{5} \big)  \notag \\
&  + \sqrt{6} (\mu_0- 1) x_{2}
       \Big( 40h + 132\mu_0^2 - 48\mu_0 
           - 12(h + \lambda_0\mu_0)x_{3}
           + 3(\mu_0 -1)x_{5}\big(8(-4\mu_0 + 1) + 5(\mu_0 - 1)x_{5}\big)
       \Big)  \notag \\
&  + 24\sqrt{6}\, h(\mu_0 - 1)x_{2}^3 \big(h +1\big)    + 12x_{2}^2 
       \Big(
           -h(15\mu_0 - 8)(\mu_0 -1)
           - (2h + 3\mu_0^2)
           + (2h + 3\mu_0) \notag \\
&
           + 3(\mu_0 - 1)x_{5}
              (3h\mu_0 - 3h + \mu_0  - 1)
       \Big)
 \Big)  + 3x_{5} 
 \Big(
     \sqrt{6}(\mu_0 -1)x_{2}(-2\mu_0 + (\mu_0 -1)x_{5})(-6\mu_0 + 4 + 3(\mu_0 - 1)x_{5}) \notag \\
&  + (\mu_0 - 1)\big(12\mu_0 - 8 + (-4\lambda_0 + 6\mu_0)x_{3} - 3(\mu_0 - 1)(2 + x_{3})x_{5}\big)\notag \\
&  + 2x_{2}^2 
       \Big(
           6h\mu_0(-\mu_0 + 1)
           - 2(2h + 3\mu_0^2)
           + 2(2h + 3\mu_0)\notag \\
& 
           + 3(\mu_0 -1)x_{5}(h\mu_0 - h+ \mu_0  -1)
       \Big)
 \Big)
\Big)\Bigg/ 
\Big(4(-1 + \mu_0)
\left(
  2\sqrt{6}\,x_2(2h + 3\mu_0^2)x_4
  + 3(1 - \mu_0)(x_4 - x_5)^2
\right)\Big)\Bigg)
\end{align}

\end{document}